\documentclass[useAMS,usenatbib]{mn2e}
\usepackage{natbib}
\usepackage{graphicx}
\usepackage{dcolumn}
\usepackage{bm}
\usepackage{MnSymbol}
\usepackage{pjournal} 

\usepackage{tikz}
\usetikzlibrary{decorations.pathreplacing}

\def\be{\begin{equation}}
\def\ee{\end{equation}}
\def\bea{\begin{eqnarray}}
\def\eea{\end{eqnarray}}
\def\ba#1\ea{\begin{align}#1\end{align}}

\def\fnl{f_\mathrm{NL}}
\def\dd{\mathrm{d}}
\def\xx{\mathbf{x}}
\def\kk{\mathbf{k}}
\def\nn{\hat{n}}
\def\lu{\ell_1}
\def\ld{\ell_2}
\def\lt{\ell_3}
\def\bl{b_{\ell_1 \ell_2 \ell_3}}
\def\mr{\mathrm}

\def\um{$\mu$m }
\def\alphas{$\alpha_\mathrm{sat}$}
\def\Mmin{$M_\mathrm{min}$}
\def\Msat{$M_\mathrm{sat}$}
\def\sigmalogm{$\sigma_\mathrm{\log M}$}

\def\Msun{M_\odot}

\begin{document}

\title[NG of the CIB anisotropies I : formalism]{
Non-Gaussianity of the Cosmic Infrared Background anisotropies~I~:  Diagrammatic formalism and application to the angular bispectrum
}

\author[F. Lacasa, A. P\'{e}nin and N. Aghanim]
{F. Lacasa$^{1}$ \thanks{E-mail: Fabien.Lacasa@ias.u-psud.fr},
A. P\'enin$^{2}$
and N. Aghanim$^{1}$\\
$^{1}$ Institut d'Astrophysique Spatiale (IAS), B\^atiment 121, F-91405 Orsay
(France)\,; Universit\'e Paris-Sud 11 and CNRS (UMR 8617) \\
$^{2}$ Aix Marseille Universit\'e, CNRS, LAM (Laboratoire d'Astrophysique de Marseille) UMR 7326, 13388, Marseille, France\\
} 

\maketitle

\begin{abstract}
We present the first halo model based description of the Cosmic Infrared Background (CIB) non-Gaussianity (NG) that is fully parametric. To this end, we introduce, for the first time, a diagrammatic method to compute high order polyspectra of the 3D galaxy density field. It allows an easy derivation and visualisation of the different terms of the polyspectrum. We apply this framework to the power spectrum and bispectrum, and we show how to project them on the celestial sphere in the purpose of the application to the CIB angular anisotropies. Furthermore, we show how to take into account the particular case of the shot noise terms in that framework. Eventually, we compute the CIB angular bispectrum at 857 GHz and study its scale and configuration dependencies, as well as its variations with the halo occupation distribution parameters. Compared to a previously proposed empirical prescription, such physically motivated model is required to describe fully the CIB anisotropies bispectrum.  Finally, we compare the CIB bispectrum with the bispectra of other signals potentially present at microwave frequencies, which hints that detection of CIB NG should be possible above 220 GHz. 

\end{abstract}


\section{Introduction}
The structuration of the large scale structures and galaxies in the Universe is a long-standing field of research in cosmology, theoretically as well as observationally. Of particular interest is the clustering of galaxies as the latter are biased tracers of the underlying dark matter field. Although perturbation theory \citep[see][for a review]{Bernardeau2002} may describe the clustering of dark matter up to mildly non-linear scales and epochs, it breaks down in the regime of highly non-linear gravitational infall and does not prescribe the behaviour of galaxies and baryonic physics with respect to dark matter. \citet{Neyman1952} pioneered the description of galaxies as distributed in clusters, which were later identified as dark matter halos as the dark matter paradigm became popular. This latter description has become a fruitful tool, assuming that galaxy properties are determined by the physical characteristics of the host halo, as dark matter simulations have become available. Indeed these simulations have permitted to prescribe the distribution of mass inside halos (a.k.a. the density profile), their abundance and spatial distribution \citep[e.g.][]{NFW1997}. Then analytic or semi-analytic models prescribing the distribution and properties of different galaxy populations may be built \citep[e.g.][]{DeLucia2007}. A common analytical tool is the halo model. In this framework, all dark matter is assumed to be bound up in halos which are populated with galaxies thanks to the halo occupation distribution (HOD). The standard HOD rules the mean number of galaxies in a halo depending on its mass \citep{Berlind2003,Kravtsov2004}. Such models have been widely used to reproduce the 2-point correlation function of optically-selected galaxies, see e.g. \cite{Tinker2010a,Coupon2012} and references therein for the most recent analyses. Most applications to date have concentrated on 2-point statistics, i.e. real-space 2-point correlation function or --auto and cross-- power spectrum of tracers. \\
One tracer of galaxies and dark matter that has been studied thanks to
the halo model is the Cosmic Infrared Background (CIB). It was first
discovered by \cite{Puget1996},  and it stems from the cumulative
emission of dusty star-forming galaxies (DSFG). The UV emission from young
stars heats up the surrounding dust which consequently reemits in the
infrared (from 8 \um~to 1 mm) with a typical greybody law. The CIB  is consequently a tracer of star formation, with lower frequencies ($\nu<220$ GHz) tracing star formation at high redshifts \citep[see e.g.][]{Penin2012}, as their emission is redshifted into the Far-IR/sub-millimeter domain. Resolutions of current instruments permit to resolve directly only a small fraction of the CIB into individual sources, in particular at far-IR frequencies ($<$857 GHz) where most of the CIB is unresolved so that sources produce brightness fluctuations generating the CIB anisotropies. The CIB fluctuations trace the clustering of the underlying DSFG and their angular power spectra have been measured, in the last few years, over a wide range of wavelengths and scales \citep{Lagache2007,Viero2009,Planck-early-CIB,Amblard2011,Thacker2012,Penin2012,Planck2013-CIB}. These measurements are usually modeled in the context of the halo model associated to a model of evolution of galaxies \citep{Cooray2010,Penin2012}. Until recently, only the power spectrum of the CIB anisotropies had been measured, however statistical information is contained in the higher order moments.\\

The hierarchy of n-point correlation functions, for n up to infinity, characterises statistically a field, univoquely under some regularity conditions. In particular beyond n=2 it probes the non-Gaussianity of the field. Non-Gaussianity (NG) studies have  emerged as a research field of interest, as they bring information complementary to power spectrum (or 2-point correlation function) analyses. They are of particular interest for the Cosmic Microwave Background (CMB), for instance, the study of primordial NG discriminates inflation models which are degenerate at the power spectrum level. Lately, the Planck NG constraints have ruled out several primordial models, in particular the possibility of ekpyrotic/cyclic Universe \citep{planck2013-NG}.\\

Nevertheless, such measurements are delicate as millimeter observations dedicated to the CMB are contaminated by foregrounds which are non-Gaussian. Extragalactic point-sources are of particular importance because they are present all over the sky and the fainter ones cannot be detected nor masked. At CMB frequencies, two types of extragalactic point-sources are present : radio-loud sources powered by an Active Galactic Nucleus \citep{Toffolatti1999} and DSFG constituting the Cosmic Infrared Background \citep{Lagache2005}. NG of these point-sources has first been looked for at radio frequencies with WMAP, focusing on radio sources which
can be considered unclustered \citep{Toffolatti1998}. Bispectrum
predictions based on number counts and measurement on WMAP data have
been found in agreement \citep{Komatsu2003} and have permitted to
quantify the level of unresolved sources in WMAP maps. At higher
frequencies, non-Gaussianity of DSFG has been pioneered by
\cite{Argueso2003,Gonzalez-Nuevo2005}, and lately \cite{Lacasa2012a} with a
phenomenological prescription based on the clustered power
spectrum. Prior to the present study, no physically-based model of the CIB NG was proposed.\\

This article builds a halo model description of galaxy clustering at high orders that we apply to predict the NG of CIB anisotropies. This allows for a full model for
the CIB anisotropies which, given a galaxy emission model and HOD
parameters, computes the power spectrum as well as the bispectrum, and
possibly higher order moments. The clustering part of the model is
fully parametric which will, at longer term, allow us to constrain these
parameters using as much statistical information from
the data as possible. In a companion article \citep{Penin2013}, referred to as Paper2 hereafter,  we carry out a Fisher analysis forecast of how the degeneracies of these parameters are broken when combining power spectra and bispectra constraints. In addition, we study in details, amongst others, the variation of the CIB angular bispectrum with respect to the models of evolution of galaxies, and the frequency evolution of the redshift-halo mass contributions to the bispectrum.

The present article is organised as follows. Section \ref{Sect:halomodel}
details the halo model formalism, accounting for the shot-noise due to
the discreteness of galaxies, the occupation statistics (HOD) and shows
the resulting 3D power spectrum. In Sect.\ref{Sect:halohighord}, we
derive the galaxy bispectrum and describe a diagrammatic method to
carry the derivation to higher orders. Sect.\ref{Sect:onsky}
introduces harmonic transform of correlation functions on the sky and shows
how the 3D polyspectra of a signal project onto the sphere. Taking the
example of the CIB, we discuss how the shot-noise terms must be accounted
for and the necessary regularisation at low redshift. The resulting CIB angular bispectrum is shown
with its different terms in Sect.\ref{Sect:results} as well as  their dependencies on the HOD parameters. We also investigate the halo-mass contributions to the galaxy bispectrum.
Eventually, in Sect.\ref{Sect:discussion}, we compare the obtained CIB
angular bispectrum to a previously proposed prescription and to the
bispectra of other signals present at microwave frequencies, namely radio sources and CMB.
We finally conclude in Sect.\ref{Sect:conclusion}.


\section{Galaxy clustering with the halo model}\label{Sect:halomodel}

The most common tool to measure the clustering of galaxies is the
two-point correlation function. At first, such measurements were well
reproduced by a simple power
law \citep{Davis1983,Madgwick2003,LeFevre2005}. Nevertheless, the progress in the field of
large scale surveys enabled more accurate measurements
that rule out such simple modeling \citep[e.g.][]{Zehavi2004} as they display a cut-off at intermediate scales.
A more complex modeling was required leading to the wide use of the
halo model \citep[e.g.][]{Cooray-Sheth2002}. In that framework, dark matter is assumed to be bound up in halos which are virialised spherical objects $\Delta_{\mathrm{vir}}$ times denser
than the background\footnote{$\Delta_{\mathrm{vir}}$ is the density
  contrast with respect to the critical density at the halo redshift.}.
The introduction of the galaxies within the halos is done through the
Halo Occupation Distribution \citep{Kravtsov2004,Tinker2010b}. This modeling has proven to be a very
convenient analytical tool to reproduce and interpret the non-linear
clustering of dark matter halos as well as that of galaxies \citep[e.g.][]{Coupon2012}. This section summarizes the ingredient of the halo model necessary to predict the galaxy distribution, and we show how to treat self-consistently the discreteness of galaxies.

\subsection{Halo framework}
In the halo model framework, galaxies reside in dark
matter halos that are assumed spherical.
Hence the galaxy density field at a given point $\xx$ reads
(redshift dependencies are implicit throughout this article, we
state them explicitly when needed)~: 
\be
\label{Eq:ngalhalobase}
n_{\mathrm{gal}}(\xx) = \sum_i n_{\mathrm{gal}}(\xx|i) 
\ee 
with $i$ being the halo index. In the literature, $n_{\mathrm{gal}}(\xx|i)$ is assumed,
implicitly or explicitly, to be a smooth distribution following the
halo density profile. However, galaxies are discrete objects\,; hence we write~: 
\be 
n_{\mathrm{gal}}(\xx|i) = \sum_{j=1}^{N_{\mathrm{gal}}(i)}
\theta(\xx-\xx_j) 
\ee 
with $j$ being the index of the random galaxies,
$N_{\mathrm{gal}}(i)$ being the --random-- number of galaxies in the
halo $i$, $x_j$ is the --random-- position of the $j$th galaxy, and
$\theta(x)$ is the galaxy profile. We assume here after that galaxies are drawn independently in the halo.\\
Equation \ref{Eq:ngalhalobase} can then be rewritten as:
\bea\label{Eq:ngalx}
\nonumber n_\mathrm{gal}(\xx) &=& \int \dd M \,\dd^3\xx_\mathrm{h} \sum_i \delta(M-M_i) \,
\delta^{(3)}(\xx_\mathrm{h}-\xx_i)\\
&&\times \int \dd^3\xx_\mathrm{g} \sum_{j=1}^{N_\mathrm{gal}(i)} 
\delta^{(3)}(\xx_\mathrm{g}-\xx_j) \,\theta(\xx-\xx_j)
\eea
with $M_i$ and $\xx_i$ the mass and position of the dark matter halo 
$i$. This equation serves as the basis for the computation of the 
galaxy clustering throughout this article.\\
Furthermore, we make the following set of assumptions for the galaxy distribution~:
\begin{itemize}
\item halos are spherical. This is a common assumption in halo models\,;
  inclusion of halo shapes was shown to have a 5-10\% effect on the 3D bispectrum by \cite{Smith2006}.
\item the number of galaxies $N_{\mathrm{gal}}(i)$
  ($=N_{\mathrm{gal}}(M_i)$) is drawn from the HOD and depends on the mass $M_i$ of the halo
  (see Sect.\ref{sect:hod}). 
\item the galaxy positions follow the dark matter halo profile.
They are drawn from a distribution whose pdf is the 
normalised halo profile $u(\xx |M)$ centered on the halo center~:
\be
p(\xx_j | i) = u(\xx_j-\xx_i | M_i)
\ee

\item the galaxy profile $\theta(x)$ is a Dirac
  $\delta^{(3)}(x)$. This assumption holds since the scales probed are
  larger than the galaxy size.
\end{itemize}
The galaxy density field may then be characterised with a hierarchy
of n-point correlation functions of $n_\mathrm{gal}$. At first 
order (n=1), the mean
number of galaxies per comoving volume, is~:
\be
\overline{n}_\mathrm{gal} = \int \dd M \,\langle N_\mathrm{gal}(M) \rangle \,\frac{\dd n_\mathrm{h}}{\dd M}
\ee
where $\frac{\dd n_\mathrm{h}}{\dd M}$ is the number of halos with mass $M$ 
per comoving volume, i.e., the halo mass function. It is convenient 
to define the galaxy density contrast as~:
\be
\delta_\mathrm{gal}(\xx) = \frac{n_\mathrm{gal}(\mathbf{x}) - 
\overline{n}_\mathrm{gal}}{\overline{n}_\mathrm{gal}}
\ee
In the following, we will derive the correlation functions of 
$\delta_\mathrm{gal}$ and their Fourier transform. To this end we need to 
explicit the behaviour of the number of galaxies occupying a halo.


\subsection{Occupation statistics}\label{sect:hod}
High resolution dissipationless simulations as well as semi-analytic
and $N$-body+gas dynamics studies show that the number of galaxies
within a single halo depends on halo mass with a shape consisting of a step, a shoulder and
a power-law tail at high mass \citep[e.g.][]{Berlind2003,Kravtsov2004}. 
This behaviour can be understood when the number of
galaxies, described as a random distribution, is split into the 
contribution from central galaxies and that
of satellite ones $N_\mathrm{gal} =
N_\mathrm{cen} + N_\mathrm{sat}$. The former is described as a
step-like function while the latter is a power law. High resolution 
simulations have brought a lot of progress in the modeling of these
two contributions \citep[e.g.][]{Tinker2010b}.

The HOD provides us with the number of galaxies in a halo\,; 
$N_\mathrm{cen}$ takes either the value of 0 or 1, and the presence of satellite 
galaxies is conditioned to $\{N_\mathrm{cen}=1\}$. If $N_\mathrm{cen}=1$, 
the number of satellites is drawn from a Poisson distribution 
\citep[see][]{Zheng2005} with mean $\overline{n}_\mathrm{sat}$ (in 
the following, over-bar denotes 
the average conditioned to $\{N_\mathrm{cen}=1\}$)\footnote{For example we have $\overline{n^2_\mathrm{sat}} = \overline{n}_\mathrm{sat}^2 + \overline{n}_\mathrm{sat}$, through the properties of the Poisson distribution.}. Hence we have :
\be
\langle N_\mathrm{sat} \rangle = P(N_\mathrm{cen}\! =\! 1) \times \overline{n}_\mathrm{sat} = \langle N_\mathrm{cen} \rangle \times \overline{n}_\mathrm{sat}
\ee
Using the properties of the Poisson distribution, we can then compute 
the expectation values that will be used in the following sections:
\bea
\nonumber \langle N_{\mathrm{gal}} \rangle &=& P(N_{\mathrm{cen}}\! =\! 1) 
\times (1+\overline{n}_{\mathrm{sat}})\\
 &=& \langle N_{\mathrm{cen}}\rangle + 
\langle N_{\mathrm{sat}}\rangle \\
\nonumber \langle N_{\mathrm{gal}}(N_{\mathrm{gal}}-1)\rangle &=& P(N_{\mathrm{cen}}\! =\! 1) 
\times \overline{(n_{\mathrm{sat}} +1) n_{\mathrm{sat}}} \\
 &=& \langle N_{\mathrm{cen}}\rangle (\overline{n}_{\mathrm{sat}}^2+2\,\overline{n}_\mathrm{sat})\\
\nonumber\langle N_\mathrm{gal} \, (N_\mathrm{gal}-1)(N_\mathrm{gal}-2)\rangle &=& P(N_\mathrm{cen}\! =\! 1) \\ \nonumber && \times \overline{(n_\mathrm{sat}+1) \, n_\mathrm{sat} \, (n_\mathrm{sat}-1)}\\
 &=& \langle N_\mathrm{cen}\rangle \left(\overline{n}_\mathrm{sat}^3+3\overline{n}_\mathrm{sat}^2\right)
\eea

We take the mean number of central galaxies as in \cite{Penin2012}:
\be
\langle N_\mathrm{cen} \rangle = P(N_\mathrm{cen}\! =\! 1) = \frac{1}{2} \left[ 1+\mathrm{erf}\left(\frac{\log M - \log M_\mathrm{min}}{\sigma_{\log M}}\right) \right]
\ee
and the mean number of satellites as:
\be
\langle N_\mathrm{sat} \rangle = \frac{1}{2} \left[ 1+
\mathrm{erf}\left(\frac{\log M - \log 2M_\mathrm{min}}{\sigma_{\log M}}\right) 
\right] \left( \frac{M}{M_\mathrm{sat}}\right)^{\alpha_\mathrm{sat}}
\ee
where, as for the rest of this article, we use base-10 logarithm.

Such expressions are motivated by hydrodynamical cosmological simulations
\citep{Berlind2003} as well as high-resolution collisionless
simulations \citep{Kravtsov2004}. In this formulation, the HOD is 
thus characterised by four parameters : $M_\mathrm{min}$,
the mass threshold above which a halo contains a central galaxy\,;
$\sigma_{\log M}$ describing the width of the transition from 0 to 1
central galaxy\,; $M_\mathrm{sat}$ the typical mass above which a halo
contains satellite galaxies\,; and $\alpha_\mathrm{sat}$ the index of
the power law for the number of satellites at large halo
masses. Furthermore, $N_\mathrm{sat}$ has a cut-off of the same 
form as the central
occupation but with a transition mass twice larger than that of the
central galaxy. This prevents halos which have a low probability of hosting a
central galaxy to contain satellite galaxies. Throughout this article we use $\log M_\mathrm{min}/\Msun = 12.6$, $\sigma_{\log M}=0.65$, $\log M_\mathrm{sat}/\Msun = 13.6$ and $\alpha_\mathrm{sat}=1.1$,
unless otherwise stated \citep[see][]{Penin2013}.


\subsection{Power spectrum}

In the last decade, it has been shown that the 2-point correlation
function of galaxies departs from a simple power law
\citep{Zehavi2004}. In the framework of the halo model, this has been
reproduced by splitting the total 2-point correlation function, or the
angular power spectrum, into two contributions, the 1- and 2-halo
terms.  The discreteness of galaxies further adds a shot-noise term,
which can be accounted for following a counts-in-cells approach
\citep{Peebles1980}. We thus have\footnote{See detailed derivation in Appendix \ref{App:derivgalpowsp}, including the shot-noise terms consistently}~:
\be
P_\mathrm{gal}(k)=P^\mathrm{1h}_\mathrm{gal}(k) + P^\mathrm{2h}_\mathrm{gal}(k) + P^\mathrm{shot}_\mathrm{gal}(k)
\ee
The shot-noise contribution is given by
\be
P^\mathrm{shot}_\mathrm{gal}(k) = \frac{1}{\overline{n}_\mathrm{gal}},
\ee
and will be examined in more detail in Sect.\ref{Sect:shot-noise}.
The 1- and 2-halo terms describe respectively the contribution of two galaxies 
within one same halo and that of two galaxies in two different halos. The 1-halo contribution is:
\be
P^\mathrm{1h}_\mathrm{gal}(k) = \int\! \dd M \,\frac{\dd n_\mathrm{h}}{\dd M}\frac{\langle N_\mathrm{gal}(M) (N_\mathrm{gal}(M)-1)\rangle}{\overline{n}_\mathrm{gal}^2} u(k|M)^2
\ee
where $u(k|M)$ is the Fourier transform of the normalised halo profile. 
Throughout this article, we use the Navarro-Frenk-White halo profile 
\citep{NFW1997}, and the Sheth \& Tormen mass function \citep{ShandT99} that provides us with associated bias functions, in particular the second order bias $b_2(M)$, introduced in
 Sect.\ref{Sect:bispectrumderiv}, which will be needed for the bispectrum computation later on.\\
The 2-halo contribution writes:
\be
\nonumber P^\mathrm{2h}_\mathrm{gal}(k) = P_\mathrm{lin}(k) \left(\int \dd M \, \frac{\langle N_\mathrm{gal}(M)\rangle \, \frac{\dd n_\mathrm{h}}{\dd M}}{\overline{n}_\mathrm{gal}} \, b_1(M) \, u(k|M) \right)^2
\ee
where $P_\mathrm{lin}(k)$ is the dark matter power spectrum 
predicted by linear theory (working at tree-level). \\

\begin{figure*}
\begin{center}
\includegraphics[width=\linewidth]{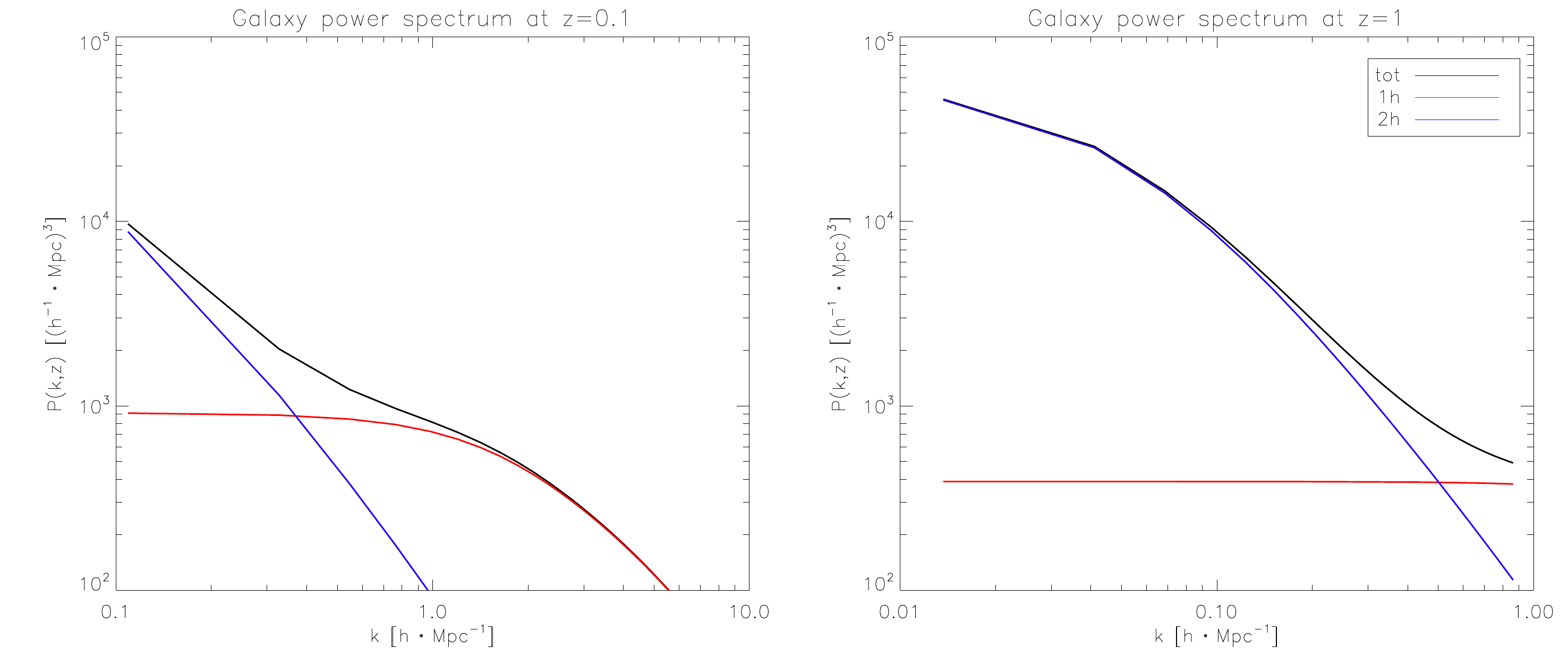}
\caption{For the Navarro-Frenk-White halo profile and the Sheth \&
  Tormen mass function we plot the 1-halo and 2-halo terms of the
  galaxy power spectrum at respectively $z$=0.1 (left panel) and $z$=1
  (right panel). Note that the $k$ range is not identical between the
  plots.  }
\label{Fig:Pk1h2h}
\end{center}
\end{figure*}

On scales larger than the typical halo size, $u(k|M) \rightarrow 1$
when $k \rightarrow 0$. Hence, the 1-halo term tends towards a constant
while the 2-halo term tends towards the linear power
spectrum. On small scales, the halo profile smoothes out both
terms. For instance, the contribution of massive halos is smoothed at a smaller cut-off wavenumber compared to that of lower mass halos. 
The 1- and 2-halo  terms of the power spectrum are exhibited in 
Fig.~\ref{Fig:Pk1h2h} at redshifts $z=0.1$ (left panel) and $z=1$
(right panel).
Note that the $k$ range is different from a panel to another. Indeed,
we compute the power spectrum at the wave-vectors which will project on
observable angular scales on the sky (see Sect.\ref{Sect:results}).
As expected, we find that the 1-halo term dominates at small scales while the
2-halo term is more important at large scales. We also note that the 1-halo
term increases clearly with time, by a factor $\sim$2, while the 2-halo does not significantly 
vary.

\begin{figure}
\begin{center}
\includegraphics[width=\linewidth]{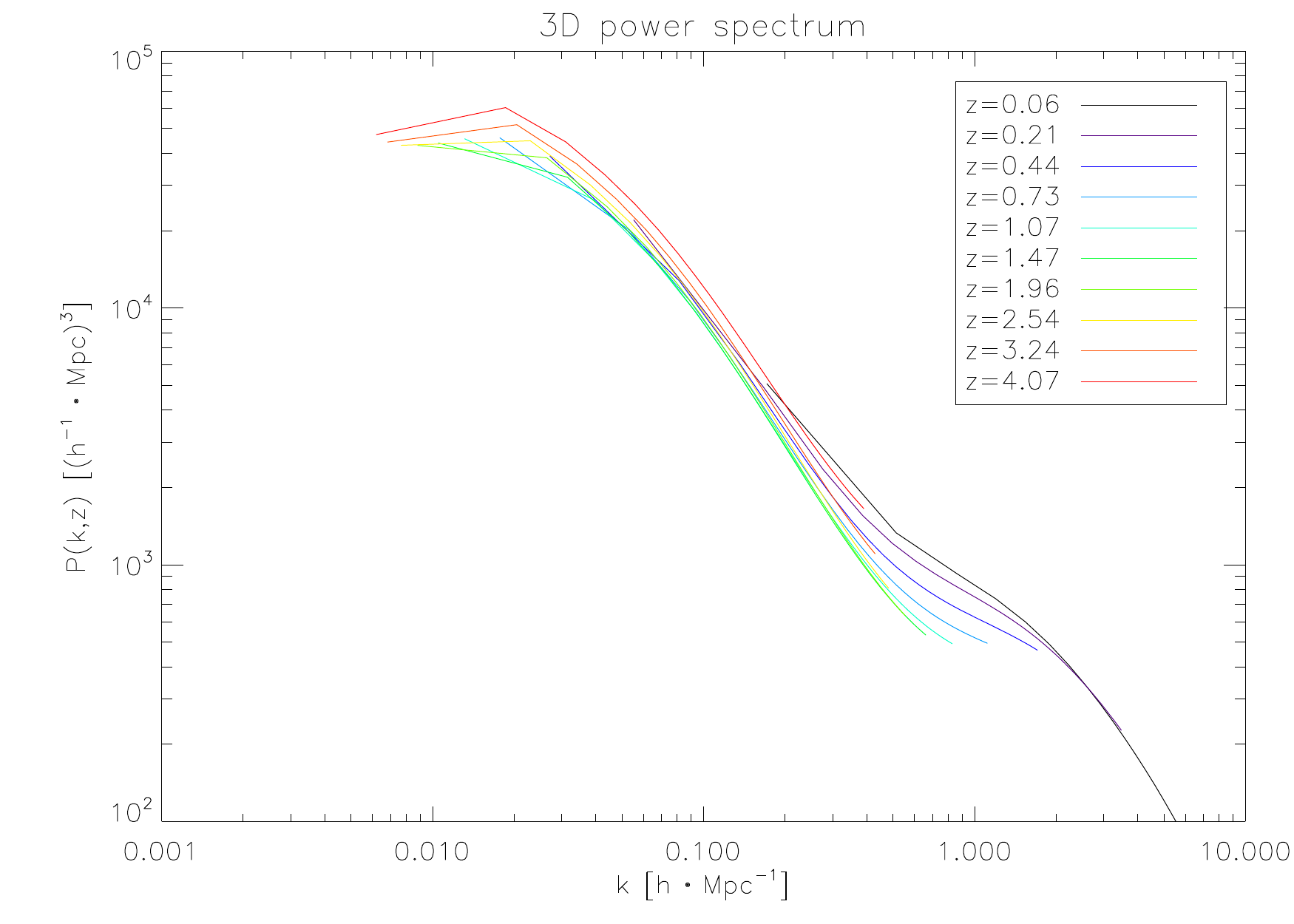}
\caption{Total 3D galaxy power spectrum as a function of redshift}
\label{Fig:Pk_z}
\end{center}
\end{figure}

In Fig. \ref{Fig:Pk_z}, we show the evolution with redshift of the
total galaxy power spectrum (1- and 2-halo terms). The spectrum
decreases with increasing redshift up to $z$$\sim$2 beyond which it
increases with redshift. This behaviour seems counter-intuitive as the
linear dark-matter power spectrum, $P_\mathrm{lin}(k,z)$, decreases
monotonically with increasing redshift. However, galaxies are biased
tracers of matter and they are strongly biased at high redshift
($b_\mathrm{gal} \sim 5.5$ at $z$ $\sim$ 4) \citep{Coupon2012,Jullo2012}, which counterbalances the decrease of
$P_\mathrm{lin}(k,z)$.


\section{Higher orders with the halo model}\label{Sect:halohighord}

As it is analytical, the halo model can be extended easily to higher orders.
Indeed, it has been already used to compute the real-space 3-point
correlation function \citep{Wang2004}, comparing favourably with simulations and measurements, as well as the bispectrum in redshift space \citep{Smith2008}, comparing again favourably with numerical simulations of dark matter.\\
In the following, we first summarize the 3D bispectrum computation and then
we present a new diagrammatic method that can be used to compute the series of high order moments.

\subsection{Bispectrum}\label{Sect:bispectrumderiv}
The 3D galaxy bispectrum at a given redshift can be written as the 
sum of several terms (see detailed derivation in Appendix 
\ref{App:derivgalbisp}, neglecting primordial non-Gaussianity 
as argued in the Appendix)~:
\begin{align}
\nonumber B_\mathrm{gal}(k_1,k_2,k_3,z) &= 
B^\mathrm{1h}_\mathrm{gal}(k_1,k_2,k_3,z)+B^\mathrm{2h}_\mathrm{gal}(k_1,k_2,k_3,z)\\
\nonumber &+B^\mathrm{3-halo}_\mathrm{gal}(k_1,k_2,k_3,z) + B^\mathrm{shot2g}_\mathrm{gal}(k_1,k_2,k_3,z) \\
&  + B^\mathrm{shot1g}_\mathrm{gal}(k_1,k_2,k_3,z)
\end{align}
where the 1-halo term writes~:
\begin{align}
\nonumber B^\mathrm{1h}_\mathrm{gal}(k_1,k_2,k_3,z) &= \int \dd M \, \frac{\langle N_\mathrm{gal} (N_\mathrm{gal}-1)(N_\mathrm{gal}-2)\rangle}{\overline{n}_\mathrm{gal}^3(z)} \, \frac{\dd n_\mathrm{h}}{\dd M} \\
&\qquad \times u(\kk_1 | M, z) \, u(\kk_2 | M, z) \, u(\kk_3 | M, z)
\end{align}
The 2-halo term writes~:
\ba
\nonumber B^\mathrm{2h}_\mathrm{gal}(k_1,k_2,k_3,z) &= \mathcal{G}_1(k_1,k_2,z) \; P_\mathrm{lin}(k_3 , z) \, \mathcal{F}_1(k_3,z) \\
\nonumber & +\ \mathcal{G}_1(k_1,k_3,z)\; P_\mathrm{lin}(k_2 , z) \,\mathcal{F}_1(k_2,z) \\
& +\ \mathcal{G}_1(k_2,k_3,z) \; P_\mathrm{lin}(k_1 ,z ) \, \mathcal{F}_1(k_1,z) \qquad
\ea
The 3-halo term writes~:
\ba
\nonumber B^\mathrm{3-halo}_\mathrm{gal}(k_1,k_2,k_3,z) &= \mathcal{F}_1(k_1,z)\, \mathcal{F}_1(k_2,z)\, \mathcal{F}_1(k_3,z)\\
\nonumber & \times \left[F^s(\mathbf{k_1},\mathbf{k_2})\, P_\mathrm{lin}(k_1,z)\, P_\mathrm{lin}(k_2,z) +\mathrm{perm.}\right] \\
\nonumber & +\ \mathcal{F}_1(k_1,z)\, \mathcal{F}_1(k_2,z)\, \mathcal{F}_2(k_3,z)\\
\label{Eq:Bk3h} & \quad \times P_\mathrm{lin}(k_1 , z)\, P_\mathrm{lin}(k_2 , z) + \mathrm{perm.}
\ea
with
\be
\mathcal{F}_1(k,z) = \int \dd M \frac{\langle N_\mathrm{gal}(M)\rangle}{\overline{n}_\mathrm{gal}(z)} \,\frac{\dd n_\mathrm{h}}{\dd M}(M,z) \, b_1(M,z) \, u(k|M,z).
\ee
where $b_1(M,z)$ is the first order bias, 
\be
\mathcal{F}_2(k,z) = \int \dd M \frac{\langle N_\mathrm{gal}(M)\rangle}{\overline{n}_\mathrm{gal}(z)} \,\frac{\dd n_\mathrm{h}}{\dd M}(M,z) \,b_2(M,z) \, u(k|M,z)
\ee
where $b_2(M,z)$ is the second order bias,
\ba
\nonumber \mathcal{G}_1(k_1,k_2,z) = {} & \int \dd M \frac{\langle N_\mathrm{gal}(N_\mathrm{gal}-1)\rangle}{\overline{n}_\mathrm{gal}(z)^2} \,\frac{\dd n_\mathrm{h}}{\dd M}(M,z) \, b_1(M,z) \\
& \qquad \times u(k_1|M,z) \, u(k_2|M,z)
\ea
and
\be\label{Eq:Fskernel}
F^s(\mathbf{k_i},\mathbf{k_j})=\frac{5}{7}+\frac{1}{2}\cos(\theta_{ij})(\frac{k_i}{k_j} +\frac{k_j}{k_i})+\frac{2}{7}\cos^2(\theta_{ij})
\ee
which stems from non-linear evolution at second order in perturbation 
theory \citep[e.g.,][]{Fry1984,Gil-Marin2012}.\\
In the following, we will note 3hcos the part of the 3-halo 
term containing the $F_{ij}$ kernel (i.e., the first term of 
Eq.\ref{Eq:Bk3h}), and we note 3h the part involving the second 
order bias (i.e., the last lines of Eq.\ref{Eq:Bk3h})~:
\be
B^\mathrm{3-halo}_\mathrm{gal}(k_{123},z) = B^\mathrm{3h}_\mathrm{gal}(k_{123},z) + B^\mathrm{3hcos}_\mathrm{gal}(k_{123},z)
\ee
Eventually, the shot-noise terms are:
\begin{align}
B^\mathrm{shot1g}_\mathrm{gal}(k_1,k_2,k_3,z) &= \frac{1}{\overline{n}_\mathrm{gal}^2(z)}\\
B^\mathrm{shot2g}_\mathrm{gal}(k_1,k_2,k_3,z) &=\frac{P^\mathrm{clust}_\mathrm{gal}(k_1)+P^\mathrm{clust}_\mathrm{gal}(k_2)+P^\mathrm{clust}_\mathrm{gal}(k_3)}{\overline{n}_\mathrm{gal}(z)}
\end{align}
with $P^\mathrm{clust}_\mathrm{gal}(k) = P^\mr{1h}_\mathrm{gal}(k) + P^\mr{2h}_\mathrm{gal}(k)$. These terms will be examined in more details in Sect.\ref{Sect:shot-noise}.

\begin{figure*}
\begin{center}
\includegraphics[width=\linewidth]{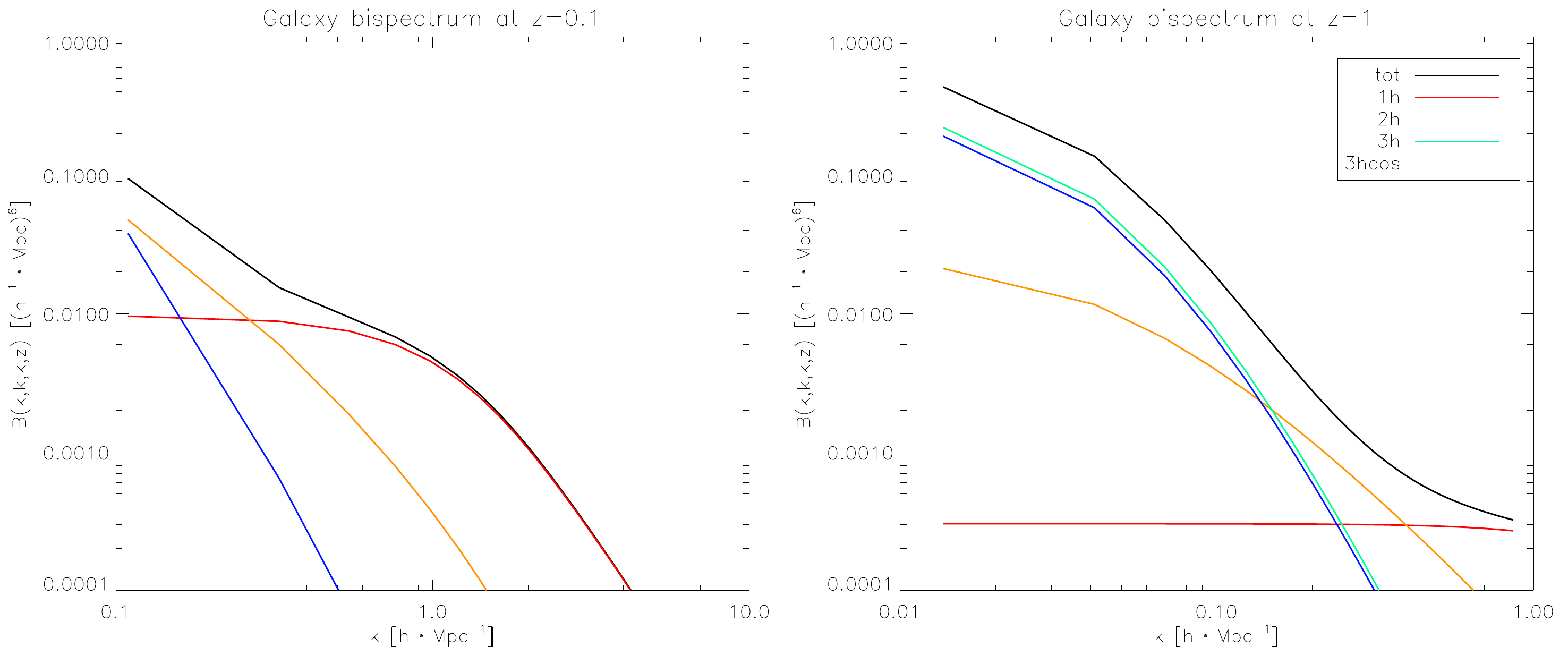}
\caption{The 1-, 2- and 3-halo terms of the galaxy bispectrum at
  respectively $z$=0.1 (left panel) and $z$=1 (right panel). Note that the $k$
  range is not identical between the plots.}
\label{Fig:Bk1h2h3h}
\end{center}
\end{figure*}

For illustration, Fig. \ref{Fig:Bk1h2h3h} shows the 1-, 2- and 3-halo
terms of the galaxy bispectrum at $z$=0.1 (left panel) and $z$=1
(right panel), in the equilateral configuration ($k_1 = k_2 =
k_3$). Note that, due to projection effects, here again the $k$ range
is not identical between the plots.

We first note that the 2- and 3-halo terms dominate at large scales
whereas the 1-halo term dominates at small scales. We see that all
terms grow with time, the 1-halo having the fastest growth. We also
note that the 3-halo term becomes negative at $z$=0.1. This comes from the
fact that the second order bias can be either positive or negative
depending on the halo mass and on the redshift. For halos with $M
\sim 10^{13-14} M_\odot$ (which typically dominate the 3-halo term)
$b_2$ is positive at high redshifts and it becomes negative at low
redshifts (e.g., at $M=10^{13} M_\odot$ it changes sign at $z \sim 0.6$,
while it changes sign at $z \sim 1.5$ at $M=10^{12} M_\odot$ and is
always positive at $M=10^{14} M_\odot$).

\begin{figure}
\begin{center}
\includegraphics[width=\linewidth]{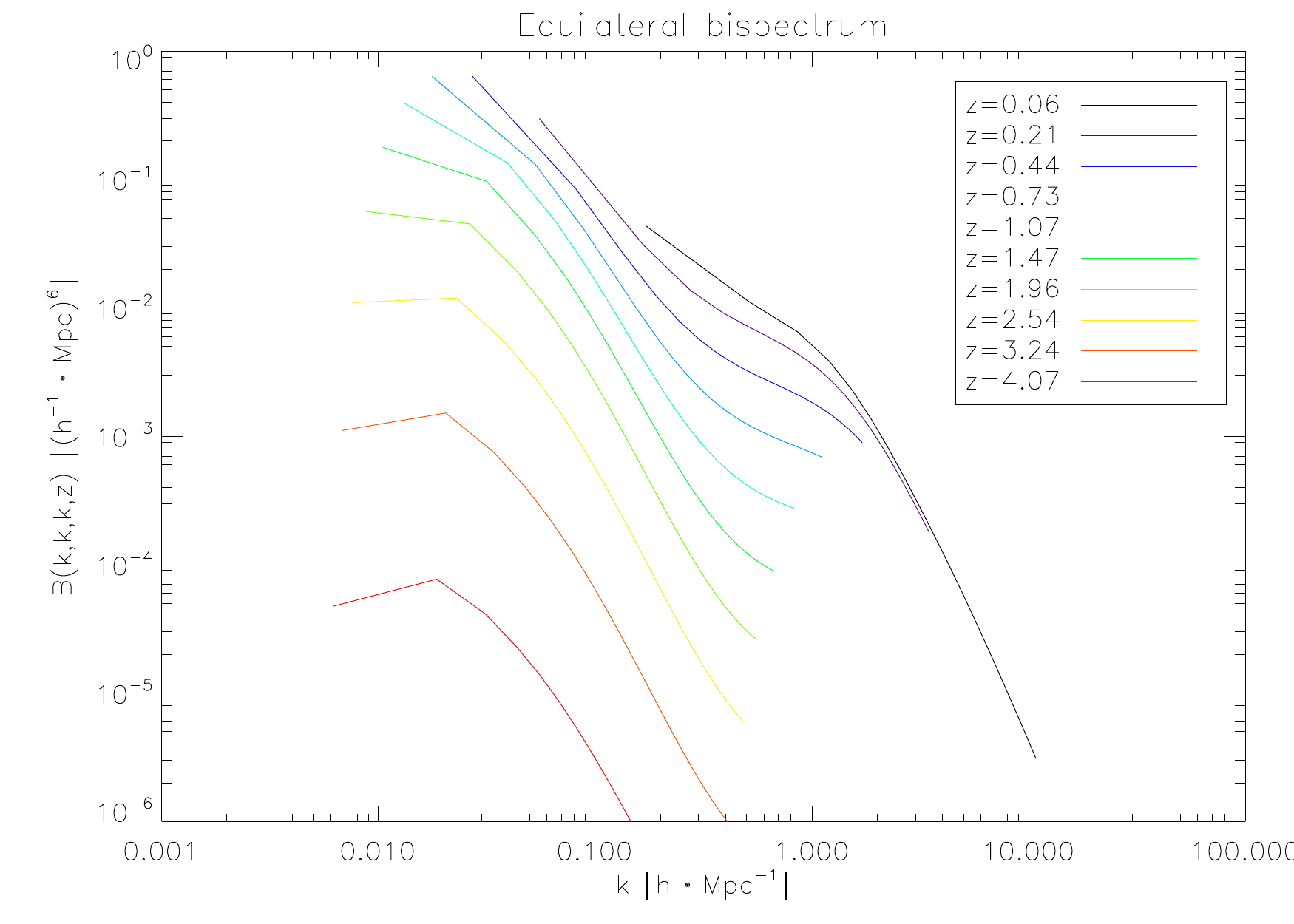}
\caption{Total galaxy equilateral bispectrum as a function of redshift.}
\label{Fig:Bk_z}
\end{center}
\end{figure}

Figure \ref{Fig:Bk_z} shows the evolution of the total 3D bispectrum
of the galaxies across redshifts. The bispectrum clearly increases
with time, as gravitational infall produces non-linear structures so
that the density field deviates more and more from its initially
Gaussian distribution.

\subsection{Diagrammatic and higher orders}\label{Sect:diagrammatic}

The 3D galaxy density field may also be characterised with higher orders than the power spectrum and bispectrum.
This is achieved through the use of polyspectra, where $\mathcal{P}^{(n)}_\mathrm{gal}(\kk_{1\cdots n} , z_{1\cdots n})$ is the polyspectrum of order $n$. See Appendix \ref{App:polyspectradef} for a comprehensive definition of polyspectra and their diagonal degrees of freedom.

We introduce here, for the first time, a diagrammatic method
permitting the derivation of the equations of the 3D galaxy polyspectrum coming
from the halo model. This approach is a generalisation of the
formalism at second and third order, the power spectrum and
bispectrum, respectively. It allows us to have a clear representation and understanding of the different terms involved. It further allows us to avoid cumbersome calculations at high order, by replacing them with diagram drawings.

\begin{figure}
\begin{center}
\includegraphics[width=\linewidth]{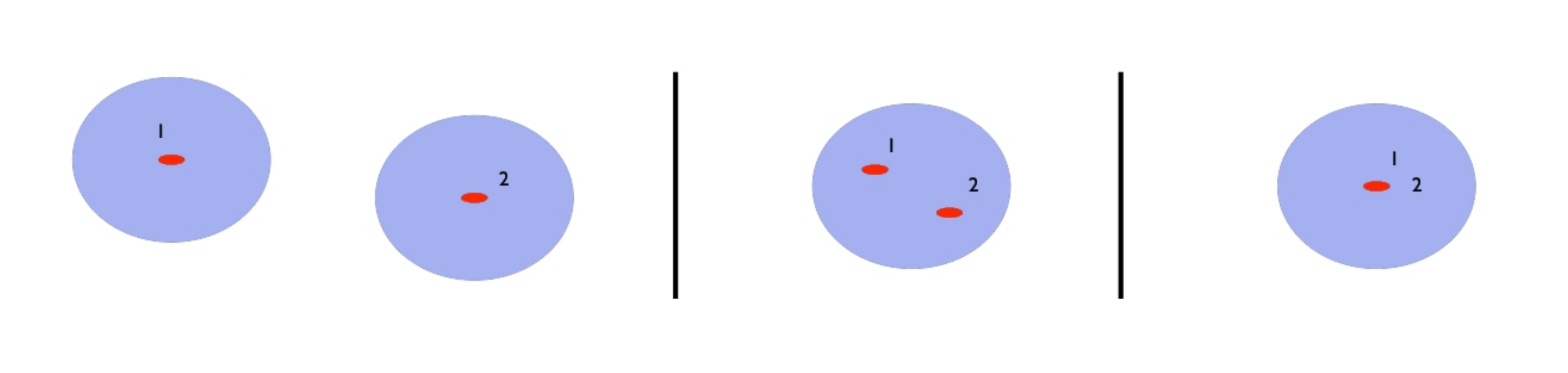}
\caption{Diagrams for the 3D galaxy power spectrum. From left to
  right: (2-halo), (1-halo,2-galaxies), (1-halo,1-galaxy). Violet
  circles represent halos and red ellipses galaxies.}
\label{Fig:spectrum_diagrams}
\end{center}
\end{figure}

\begin{figure}
\begin{center}
\includegraphics[width=\linewidth]{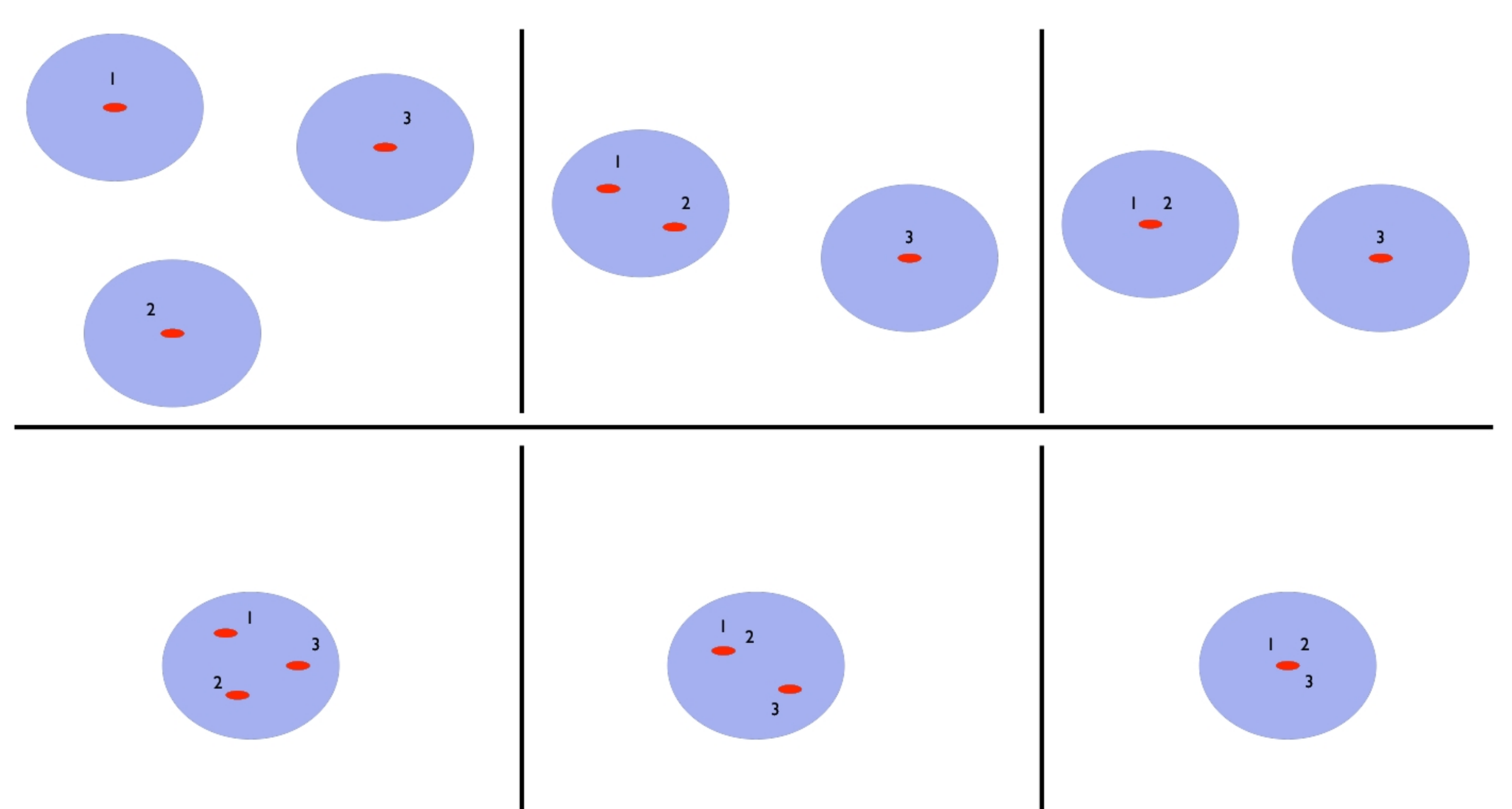}
\caption{Diagrams for the 3D galaxy bispectrum. From left to right and
  top to bottom : (3-halo), (2-halo 3-galaxies), (2-halo 2-galaxies), (1-halo
  3-galaxies), (1-halo 2-galaxies), (1-halo 3-galaxies).}
\label{Fig:bispectrum_diagrams}
\end{center}
\end{figure}

\begin{figure}
\begin{center}
\includegraphics[width=\linewidth]{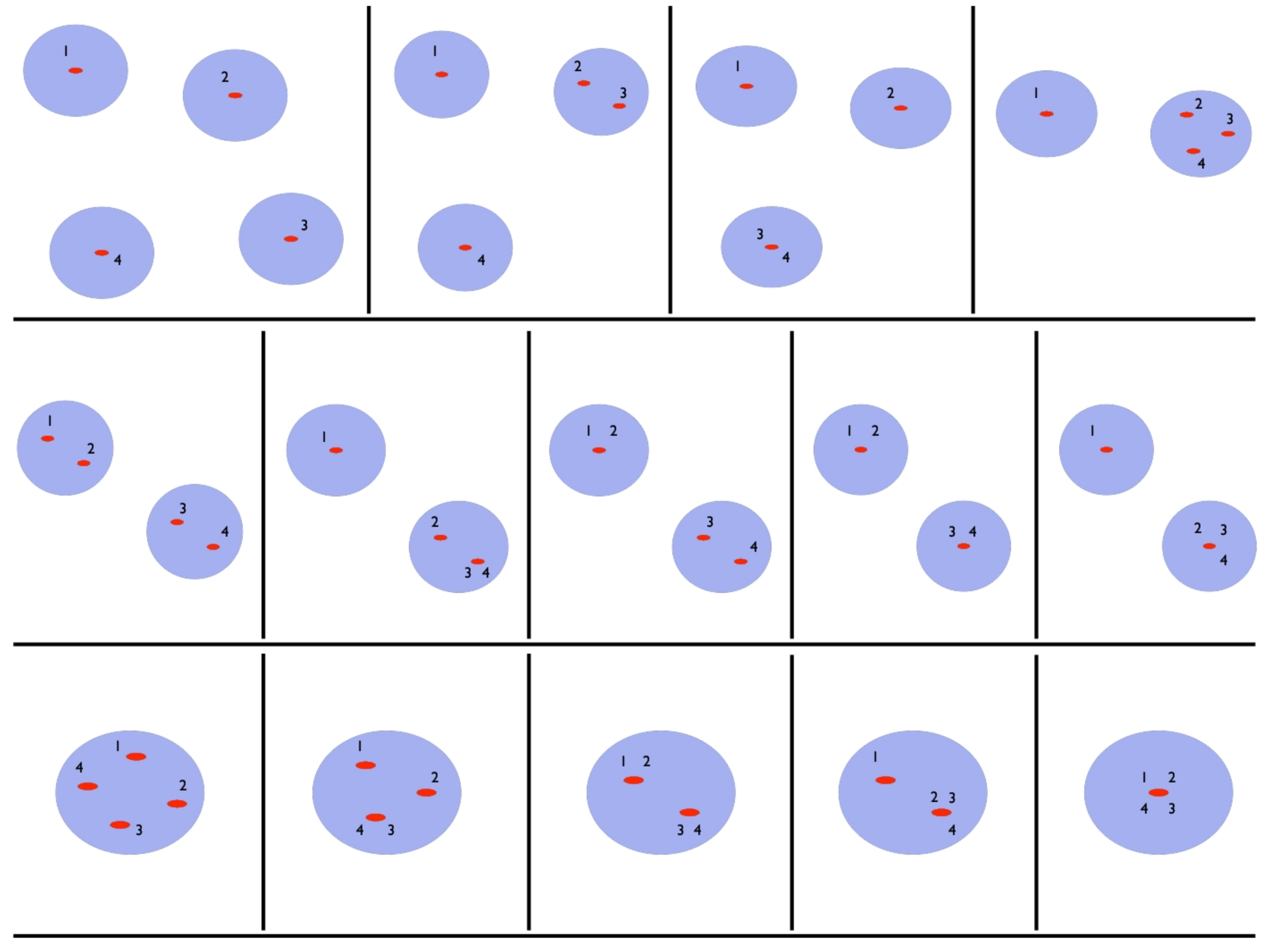}
\caption{Diagrams for the 3D galaxy trispectrum}
\label{Fig:trispectrum_diagrams}
\end{center}
\end{figure}

The first step of the diagrammatic method that we propose here is to draw in the form of diagrams all the possibilities of putting $n$ galaxies in halo(s). Potentially, two or more galaxies can lie at the same point (``contracted") for the shot-noise terms. Then for each diagram, the galaxies are labeled, e.g., from 1 to $n$, as well as the halos e.g. with $\alpha_1$ to $\alpha_p$. An illustration is given in Fig.~\ref{Fig:spectrum_diagrams} that displays the three diagrams for the power spectrum\,; Fig.\ref{Fig:bispectrum_diagrams}
shows the six diagrams for the bispectrum\,; and finally Fig.\ref{Fig:trispectrum_diagrams} exhibits the 14 diagrams for the trispectrum. 

Each diagram produces a polyspectrum term. This term contains a prefactor $1/\overline{n}_\mathrm{gal}^{\,n}$ multiplied by an integral over the halo masses $\int \dd M_{\alpha_1 \cdots \alpha_p}$ of several factors. The following ``Feynman"-like rules prescribe these different factors~:
\begin{itemize}
\item for each halo $\alpha_j$ there is a corresponding~:
 \begin{itemize}
 \item halo mass function $\left.\frac{\dd n_\mathrm{h}}{\dd M}\right|_{M_{\alpha_j}}$
 \item average of the number of galaxy uplets in that halo.\\
 e.g., $\langle N_\mathrm{gal} \rangle$ for a single galaxy in that halo, $\langle N(N-1)\rangle$ for a pair etc.
 \item as many halo profiles $u(k | M_{\alpha_j})$ as \emph{different} points, where $k=|\sum_{i\in\mathrm{point}} \kk_i|$.\\
For example $k=k_i$ for a non-contracted galaxy $i$, while $k=|\kk_{i_1} + \cdots + \kk_{i_q}|$ for a galaxy contracted $q$ times with labels $i_1 \cdots i_q$.
 \end{itemize}
 \item the final factor is the halo polyspectrum of order $p$, conditioned to the masses of the corresponding halos : 
 $$\mathcal{P}_\mathrm{halo}^{(p)}\left(\sum_{i \in \alpha_1} \kk_i \, , \cdots , \sum_{i \in \alpha_p} \kk_i \, | \, M_{\alpha_1} \, , \cdots , M_{\alpha_p} \right)$$
where the sum $\sum_{i \in \alpha_j} \kk_i$ runs over the indexes $i$ of the galaxies inside the halo $\alpha_j$.
\end{itemize}

Finally the possible permutations of the galaxy labels 1 to $n$ in the diagram are taken into account : the contribution is the sum over permutations of $\{1\cdots n\}$ which produce different diagrams. For example, we have seen in Sect.\ref{Sect:bispectrumderiv} that some contributions to the galaxy bispectrum (namely 1-halo, 3-halo and shot1g) have a single term while others (namely 2-halo and shot2g) have 3 terms.
\newline

As an example, the (2-halo, 2-galaxy) term of the bispectrum 
(upper right diagram in Fig.\ref{Fig:bispectrum_diagrams}) yields~:
\begin{align}
\label{Eq:B2h2gdiag}
\nonumber B^\mathrm{2h2g}(k_{123}) &= \frac{1}{\overline{n}_\mathrm{gal}^3} \int \dd M_{\alpha_{12}} \, \langle N_\mathrm{gal}(M_{\alpha_1}) \rangle \, \langle N_\mathrm{gal}(M_{\alpha_2}) \rangle \\
\nonumber & \times \left.\frac{\dd n_\mathrm{h}}{\dd M}\right|_{M_{\alpha_1}} \!\! \left.\frac{\dd n_\mathrm{h}}{\dd M}\right|_{M_{\alpha_2}} \!\! u(|\kk_1+\kk_2| | M_{\alpha_1}) \, u(k_3 | M_{\alpha_2}) \\
& \times \mathcal{P}^{(2)}_\mathrm{halo}(\kk_1+\kk_2, \kk_3 | M_{\alpha_1} , M_{\alpha_2}) + \mathrm{perm.}
\end{align}
Furthermore, the $\mathcal{P}^{(2)}_\mathrm{halo}$ term simplifies into $P_\mathrm{halo}(k_3 | M_{\alpha_1} , M_{\alpha_2})$ as $\kk_1+\kk_2 = -\kk_3$.

We described in the previous sections the mass function, halo profile and the HOD governing the number of galaxies in a halo. The final element needed for the computation of the galaxy polyspectra is a description of the halo polyspectra. To this end, we adopt the local biasing scheme which allows to compute the halo polyspectrum from the matter polyspectrum as described in Appendix \ref{App:derivhalo3ptcf}.

At high order, the halo polyspectrum has thus several possible sources. The first source is the first order biasing of the corresponding dark matter polyspectrum, either primordial (primordial non-Gaussianity) or from perturbation theory. In the bispectrum case, it produces the 3hcos term. The second source is the higher order biasing of lower order dark matter polyspectrum. In the bispectrum case, it produces the 3h term.

Hence, we have proposed a diagrammatic method to compute galaxy polyspectra which gives the power and simplicity of drawings to compute otherwise cumbersome equations at high order. The focus of the next section is to relate these 3D polyspectra to observables on the sky.

\section{CIB angular polyspectra on the sky}\label{Sect:onsky}

Measurements of the CIB clustering are carried out on the celestial sphere. Hence a statistical characterisation of random fields on the
sphere is needed, as well as the projection of the statistics of 3D random fields onto the sphere.\\
In this section, we first describe the formalism of correlation functions on the sphere, then we derive the CIB angular polyspectrum. Eventually, we discuss the shot noise terms and the effect of the flux cut.

\subsection{Correlation functions in harmonic space}\label{Sect:polyspsphere}

Given a full-sky map of the temperature $T(\nn)$ of some
signal on the celestial sphere, it can be decomposed in the harmonic basis as
\be
T(\nn) = \sum_{\ell m} a_{\ell m} \, Y_{\ell m}(\nn)
\ee
with
\begin{equation}
a_{\ell m} = \int \dd^2\nn \; Y^*_{\ell m}(\nn) \; \Delta
T(\nn)
\end{equation}
with the usual orthonormal spherical harmonics $Y_{\ell m}$.\\ For a
--statistically isotropic-- Gaussian field, all the statistical
information is contained in the power spectrum $C_\ell$, the 2-point
correlation function in harmonic space:
$\langle a_{\ell m} \, a^*_{\ell' m'} \rangle = C_\ell \, \delta_{\ell \ell'} 
\, \delta_{m m'}$. For non-Gaussian fields, information is also 
contained in higher-order moments. For instance, the bispectrum $\bl$ is:
\be\label{Eq:defbisp}
\langle a_{\ell_1 m_1} a_{\ell_2 m_2} a_{\ell_3 m_3} \rangle = G_{\lu \ld \lt}^{m_1 m_2 m_3} \times \bl
\ee
with the Gaunt coefficient
\begin{align}
\label{Eq:defGaunt}
G_{1,2,3} &= \int \dd^2\nn \;Y_{123}(\nn) \\
\label{Eq:GauntWigner}
&= \sqrt{\frac{(2\ell+1)_{123}}{4\pi}} 
\left(\begin{array}{ccc}
\ell_1 & \ell_2 & \ell_3 \\ 0 & 0 & 0
\end{array}\right)
\left(\begin{array}{ccc}
  \ell_1 & \ell_2 & \ell_3\\
  m_1 & m_2 & m_3
\end{array}\right)
\end{align}
where $Y_i=Y_{\ell_i m_i}$. In the following, the subscript 123 
denotes the product of the corresponding variables, e.g., 
$X_{123} \equiv X_1 \, X_2 \, X_3$. The Gaunt coefficient is zero 
unless the triplet $(\ell_1,\ell_2,\ell_3)$ follows the triangle 
inequalities and $m_1+m_2+m_3=0$.

Higher order polyspectra $\mathcal{P}^{(n)}(\ell_{1\cdots n}\, , \ell^\mathrm{d}_{1\cdots (n-3)})$ are defined in Appendix \ref{App:polyspectradef}, along with the case of diagonal independence.

The one-to-one correspondence between the full pdf and the hierarchy
of polyspectra for $n$ up to infinity
ensures that the polyspectra provide us with a full statistical
characterisation of a given field on the sphere.


\subsection{Anisotropy projection on the sky}\label{Sect:3Dto2Dproj}

The CIB is mostly unresolved in the far-infrared domain which leads to the loss of the redshift information. The emission is thus integrated on a large range of redshift ($1<z<4$), and the CIB temperature in a given direction $\nn$, is a line-of-sight integral of the IR emissivity \texttt{j}$_\nu$  per comoving volume:

\be
T(\nn,\nu) = \int \dd z \, \frac{\dd r}{\dd z} \, a(z) \, 
\texttt{j}_\nu\left(r(z)\,\nn,z\right)
\ee
with \texttt{j}$_\nu$ in Jy/Mpc so that $T$ has units of Jy/sr and may be converted 
to a temperature elevation at CMB frequencies through Planck's law. 
Here $r$ is the comoving distance to redshift $z$ and $a$ is the scale 
factor.\\
Using the Rayleigh/plane wave expansion and the Fourier expansion of 
the emissivity field, we obtain:
\be\label{eq:almirgen}
a_{\ell m}(\nu) = i^\ell \int \frac{\dd^3\kk}{2\pi^2} \,\dd z \,\frac{\dd r}{\dd z} \, a(z) \,  j_\ell(kr)\, Y^*_{\ell m}(\hat{k})\, \texttt{j}_\nu(\kk,z)
\ee
where $j_\ell$ is the spherical Bessel function of order $\ell$.

We relate in Appendix \ref{App:derivpolyspecirgen} the CIB angular polyspectrum 
to the 3D emissivity polyspectra, for the terms which are diagonal-independent\footnote{Note that not all terms may be diagonal-independent, in which case Eq.\ref{Eq:proj3Dpolysp-general} must be used in all generality.}. This gives:
\begin{align}
\nonumber \mathcal{P}^{(n)}_\mathrm{CIB}(\ell_1 , \cdots , \ell_n) &= \left(\frac{2}{\pi}\right)^n \!\!\int\! k^2_{1 \cdots n} \dd k_{1 \cdots n} \, \dd z_{1\cdots n} \, x^2 \dd x \\
\nonumber & \left[a(z_i) \left.\frac{\dd r}{\dd z}\right|_{z_i} \, j_{\ell_i}(k_i r_i) j_{\ell_i}(k_i x) \right]_{i=1\cdots n} \\ 
& \qquad \times \mathcal{P}^{(n)}_\texttt{j}(k_{1 \cdots n},z_{1 \cdots n})
\label{eq:cibpoly}
\end{align}
with parity invariance imposing $\ell_1+\cdots+\ell_n$ to be even.\\
In the Limber's approximation, Eq.\ref{eq:cibpoly} simplifies to~:
\be
\mathcal{P}^{(n)}_\mathrm{CIB}(\ell_1 , \cdots , \ell_n) = 
\int \frac{r^2 \dd r}{r^{2n}} a^n(z) \, \mathcal{P}^{(n)}_\texttt{j}(k^*_{1 \cdots n},z)
\label{eq:cibpoly_spl}
\ee
with $k^*=(\ell+1/2)/r(z)$. Note that, as a consequence of Limber's 
approximation, Eq.\ref{eq:cibpoly_spl} now involves 
the emissivity polyspectrum at a single redshift 
(i.e., $z_1 = \cdots = z_n = z$).

Galaxy power spectrum and galaxy emissivity are often related assuming implicitly or explicitly that 
all galaxies at a given redshift $z$ have the same luminosity \citep[e.g.,][]{Knox2001,Penin2012,Xia2012}. We will refer to this approach as the constant-luminosity assumption. It yields :
\be
\delta \texttt{j}_\nu(\kk,z) = \overline{\texttt{j}}_\nu(z) \cdot \delta_\mr{gal}(\kk,z)
\ee
where the average emissivity $\overline{\texttt{j}}_\nu$ is computed from galaxy 
number counts.\\
In the present study, we propose a more general assumption referred to as the flux-abundance independence assumption.
That is, we assume that the flux is stochastic with a distribution given by the number counts, and furthermore that the stochasticity is independent of the galaxy position/abundance. The commonly used constant-luminosity assumption is a special case of the flux-abundance independence assumption when the luminosity function $\frac{\dd^2 N}{\dd L \, \dd V}$ is a dirac. Note that, although the flux-abundance independence is a more general assumption than the constant luminosity one, it does not capture the possibility that the luminosity may depend on underlying variables having an influence on galaxy abundance (e.g., the host halo mass).

For non shot-noise terms, the flux-abundance independence assumption yields the same relation 
as the constant-luminosity assumption:
\be
\mathcal{P}^{(n)}_\texttt{j}(k^*_{1 \cdots n},z) = 
\overline{\texttt{j}}_\nu(z)^n \ \mathcal{P}^{(n)}_\mathrm{gal}(k^*_{1 \cdots n},z).
\ee
The difference between the two assumptions impacts the shot-noise 
terms as discussed in detail in Sect.\ref{Sect:shot-noise}.

For non shot-noise terms, the angular polyspectrum then takes the form~:
\be\label{Eq:polyspecirgen}
\mathcal{P}^{(n)}_\mathrm{CIB}(\ell_1 , \cdots , \ell_n) = \int \frac{\dd z}{r^{2n-2}} \, \frac{\dd r}{\dd z} \, a^n(z) \, \overline{\texttt{j}}_\nu(z)^n \, \mathcal{P}^{(n)}_\mathrm{gal}(k^*_{1 \cdots n},z)
\ee
In particular at second order the power spectrum is
\be\label{Eq:specirgen}
C_\ell(\nu) = \int \frac{\dd z}{r^2}\, \frac{\dd r}{\dd z}\, a^2(z) \,\overline{\texttt{j}}_\nu^{\,2}(z)\, P_\mathrm{gal}(k^*,z)
\ee
as given e.g. by \cite{Knox2001,Penin2012,Xia2012}\\
And at third order, the bispectrum is
\be\label{Eq:bispecirgen}
 \bl = \int \frac{\dd z}{r^4} \,\frac{\dd r}{\dd z}\, a^3(z) 
\,\overline{\texttt{j}}_\nu^3(z)\, B_\mathrm{gal}(k^*_{123},z)
\ee

\subsection{Shot-noise}\label{Sect:shot-noise}
The shot-noise terms, for the power spectrum or for the bispectrum, 
correspond to terms in the correlation function involving multiple 
times the same galaxy. 
For the power spectrum, the halo model gives the galaxy shot-noise 
power spectrum :
\be
P_\mathrm{gal}^\mathrm{shot}(k,z) = \frac{1}{\overline{n}_\mathrm{gal}(z)}
\ee
With the constant-emissivity assumption, we get the angular power spectrum :
\be\label{Eq:clshotem}
C_\ell^\mathrm{shot} = \int_{z=0}^\infty \frac{\dd z}{r^2} \, 
\frac{\dd r}{\dd z} \, a^2(z) \,  \overline{\texttt{j}}_\nu^2(z) \, P_\mathrm{shot}(k,z)
\ee
The shot-noise level can be predicted from number counts \citep[see e.g.,][]{Lacasa2012a} as :
\be\label{Eq:clshotdnds}
C_\ell^\mathrm{shot} = \int_{z=0}^\infty\int_0^{S_\mathrm{cut}} S^2 \, \frac{\dd^2 n}{\dd S \, \dd z} \, \dd S \, \dd z
\ee
These two formula agree if indeed all sources have the same luminosity, 
potentially depending on redshift. However in reality, sources do not 
have the same luminosity\,; and specifically Eq.\ref{Eq:clshotdnds} 
will give more weight to bright galaxies --averaging $S^2$-- as 
compared to Eq.\ref{Eq:clshotem} --averaging $S$.\\
With the flux-abundance independence assumption, the distribution of
luminosities can be incorporated in the model by introducing the 
nth-order emissivities~:
\be\label{Eq:norder-em}
\texttt{j}_\nu^{(n)}(z) = \frac{(1+z)^n \; r(z)^{2n-2}}{\frac{\dd r}{\dd z}} \int_0^{S_\mathrm{cut}} S^n \frac{\dd^2 N}{\dd S \,\dd z} \dd S
\ee
which effectively average $S^n$ instead of $S$. Removing 
$\overline{n}_\mathrm{gal}(z)$ factors from the shot-noise 3D power spectrum, 
the shot-noise angular power spectrum becomes :
\be\label{Eq:clshot1gj2}
C_\ell^\mathrm{shot} = \int \frac{\dd z}{r^2} \, \frac{\dd r}{\dd z} \, a^2(z) \,  \texttt{j}_\nu^{(2)}(z) \times 1
\ee
which can be shown to be equivalent to Eq.\ref{Eq:clshotdnds}.

At any order, with the diagrammatic approach described 
in Sect.\ref{Sect:diagrammatic}, the shot noise terms can be computed and integrated over 
redshifts with Eq.\ref{Eq:polyspecirgen} provided the following 
modification~:\\
For each contraction, $\texttt{j}_\nu^p(z)/\overline{n}_\mathrm{gal}^{\,p-1}$ must be 
replaced by the $p$-th order emissivity $\texttt{j}_\nu^{(p)}(z)$, where $p$ is the order
 of the contraction under consideration.

For example for the bispectrum, the 3D shot-noise contains three terms 
(see Appendix \ref{App:derivgalbisp})~:
\be
B^\mathrm{shot}_\mathrm{gal}(k_1,k_2,k_3,z) = B^\mathrm{1h-1g}_\mathrm{gal}+B^\mathrm{1h-2g}_\mathrm{gal}+B^\mathrm{2h-2g}_\mathrm{gal}
\ee
with
\begin{align}
B^\mr{1h-1g}_\mathrm{gal}(k_1,k_2,k_3,z) &= \frac{1}{\overline{n}_\mathrm{gal}^2(z)}\\
B^\mr{1h-2g}_\mathrm{gal}(k_1,k_2,k_3,z) &=\frac{P^\mr{1h-2g}_\mathrm{gal}(k_1)+P^\mr{1h-2g}_\mathrm{gal}(k_2)+P^\mr{1h-2g}_\mathrm{gal}(k_3)}{\overline{n}_\mathrm{gal}(z)}\\
B^\mr{2h-2g}_\mathrm{gal}(k_1,k_2,k_3,z) &=\frac{P^\mr{2h-2g}_\mathrm{gal}(k_1)+P^\mr{2h-2g}_\mathrm{gal}(k_2)+P^\mr{2h-2g}_\mathrm{gal}(k_3)}{\overline{n}_\mathrm{gal}(z)}
\end{align}
Now taking into account luminosity distribution, the 1-galaxy angular shot-noise 
takes the form :
\be\label{Eq:blshot1gj3}
b_{\ell_{123}}^\mathrm{shot1g} = \int \frac{\dd z}{r^4} \, \frac{\dd r}{\dd z} \, a^3(z) \,  \texttt{j}_\nu^{(3)}(z) \times 1
\ee
And the 2-galaxy shot-noise :
\begin{align}
\nonumber b_{\ell_{123}}^\mathrm{shot2g} &= \int \frac{\dd z}{r^4} \, \frac{\dd r}{\dd z} \, a^3(z) \,  \texttt{j}_\nu^{(1)}(z) \; \texttt{j}_\nu^{(2)}(z)\\
& \times \left[ P^\mathrm{clust}_\mathrm{gal}(k^*_1)+P^\mathrm{clust}_\mathrm{gal}(k^*_2)+P^\mathrm{clust}_\mathrm{gal}(k^*_3) \right]
\end{align}
with $P^\mathrm{clust}_\mathrm{gal}(k)=P^\mathrm{1h}_\mathrm{gal}(k,z)+P^\mathrm{2h}_\mathrm{gal}(k,z)$, and as usual $k^*~=~\frac{\ell+1/2}{r(z)}$.

In this formulation, the flux cut, or more generally the selection
function, is implemented in the $p$-order emissivities. The shot-noise
terms are quite sensitive to the value of this flux limit, which in
turn depends on the instrumental setup used for the observation. Hence,
the amplitude of the shot-noise terms may change depending on the
instrument considered.


\subsection{Flux cut and low redshift contribution}\label{Sect:reglowz}
As can be seen on Eq.\ref{Eq:clshot1gj2} and \ref{Eq:blshot1gj3}, the
shot-noise equations diverge if the emissivities tend to a constant as
$z\rightarrow 0$. This is indeed the case in the Euclidean case
($\frac{\dd n}{\dd S} \propto S^{-5/2}$) where Eq.\ref{Eq:clshotdnds}
diverges if no flux cut is applied.

In practice, the flux cut reduces the contribution from low redshift sources
which dominate the counts. This needs to be reflected in the halo
model, where the number of objects is dictated by the HOD. For
simplicity, we implement the flux cut in terms of a cut-off in the redshift
integrals at $z_\mathrm{cut}$. Below this redshift, a typical galaxy
with luminosity $L_*$ (the knee of the luminosity function) has a flux
$S \ge S_\mathrm{cut}$~:

\be
S_\mathrm{cut} = \frac{L_*}{4\pi \, d^2_L(z_\mathrm{cut})}
\ee
with $d_L$ the luminosity distance.\\
The effect of the flux cut on the galaxy clustering has not been considered previously in the CIB power 
spectrum literature, although it is theoretically necessary.
We found that it has a small effect on the power spectrum, mostly on the 1-halo term and for some values of the HOD parameters ($\alpha > 1$). The redshift cut has more effect on the bispectrum and potentially at higher order, as they may be more sensitive to low redshift because of the $r(z)^{2n-2}$ denominator in Eq.\ref{Eq:polyspecirgen} which goes to zero as $z\rightarrow0$.


\section{Results}\label{Sect:results}


\subsection{The CIB angular bispectrum}

In the following, we apply our formalism to the computation of the 
bispectrum of the CIB. To this end, the HOD best-fit parameters were obtained so as to
reproduce CIB power spectrum constraints \citep{Penin2013}. We also
use the galaxy emission model by \cite{Bethermin2011} at 350 \um = 857 GHz  with a flux cut at 0.82 Jy \citep{Planck-ERCSC}, which implies
a redshift cut at $z_\mathrm{cut} = 0.03$.\\

\begin{figure*}
\begin{center}
\includegraphics[width=0.9\linewidth]{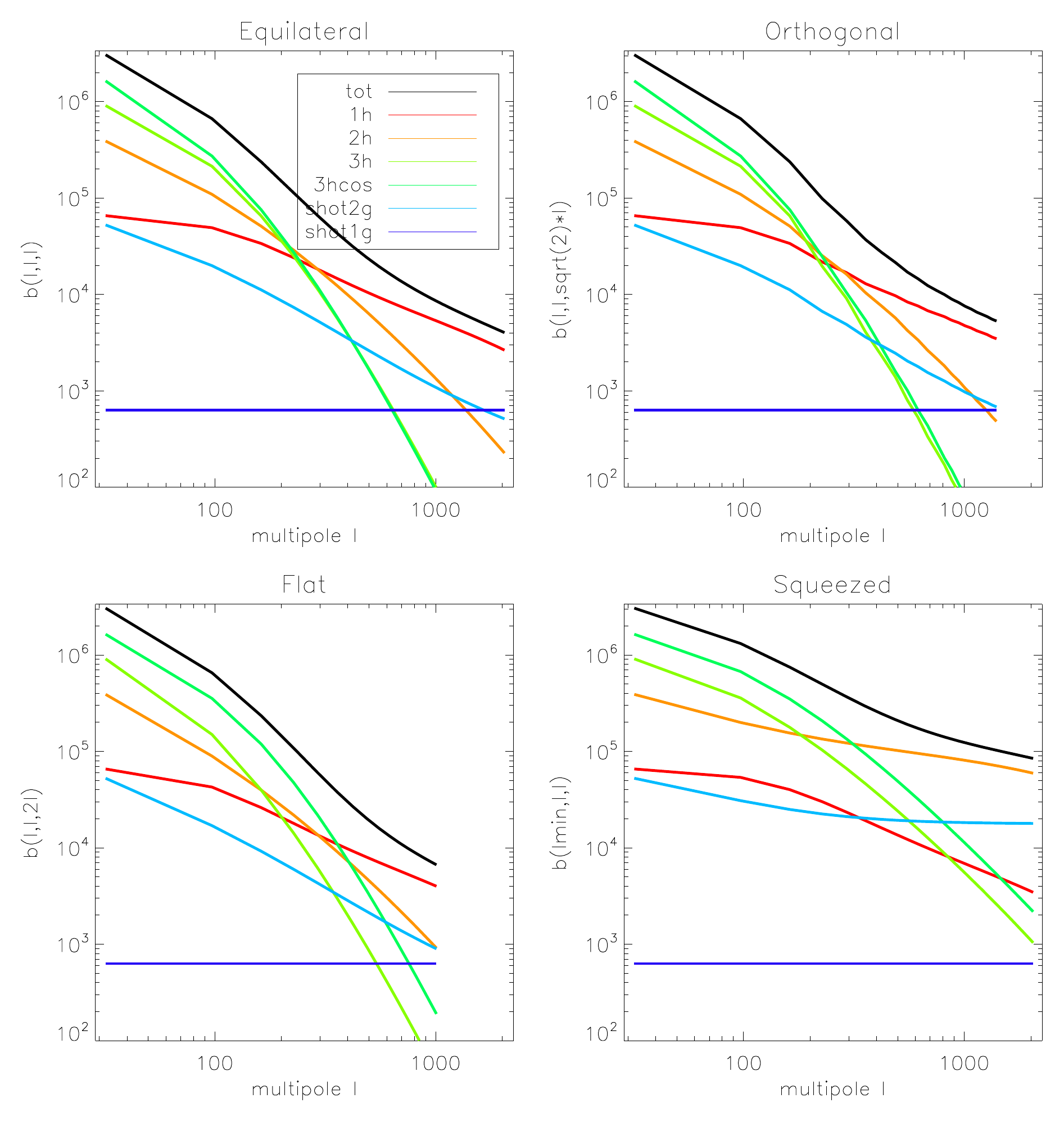}
\caption{CIB bispectrum and its different terms in some particular configurations at 350 $\mu$m = 857 GHz. The squeezed configuration uses $\ell_\mr{min} = 32$.}
\label{Fig:bl_speconf}
\end{center}
\end{figure*}

The obtained CIB bispectrum and its different terms are displayed in
Fig.\ref{Fig:bl_speconf} in some particular configurations :
equilateral, orthogonal isoceles, flat isoceles and squeezed. We
consider a multipole range of $\ell=30-2000$. On this range of scales,
shot-noise terms are found to be negligible, as happens for the power
spectrum \citep{Penin2012}. However, they are expected to dominate on
smaller scales ($\ell \sim 5000$ for the particular galaxy emission
model used in the present study).
The 2-halo and 3-halo terms dominate on large scales $\ell \leq 300$,
while the 1-halo dominates on small scales, except in the squeezed
configuration where the 2-halo term dominates for $\ell>300$.

\begin{figure*}
\begin{center}
\includegraphics[width=\linewidth]{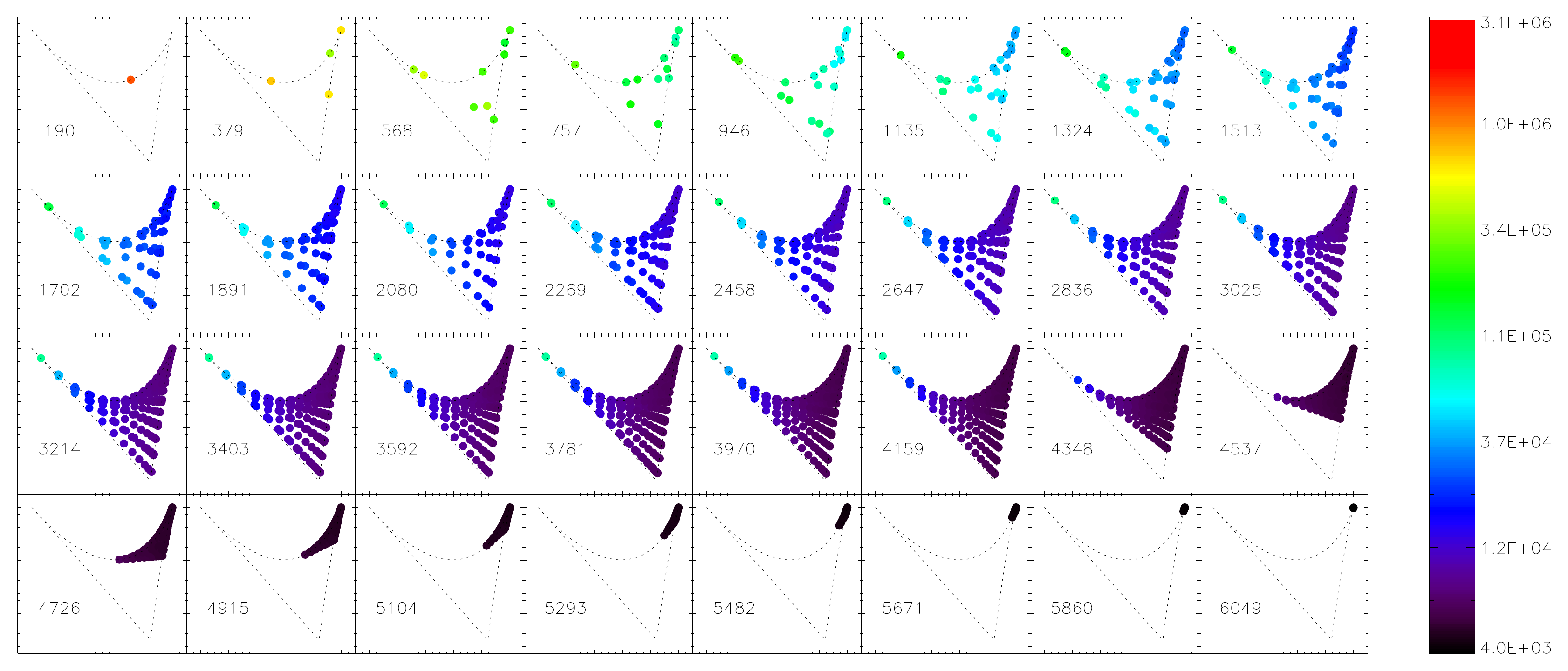}
\caption{CIB bispectrum in the geometrical parametrisation at 350 $\mu$m = 857
  GHz. It exhibits a peak in the squeezed configurations (upper left
  corner of each subplot).}
\label{Fig:bltot_inparam}
\end{center}
\end{figure*}

An interesting point is the configuration dependence of the
bispectrum. Figure \ref{Fig:bltot_inparam} shows the CIB bispectrum at
350 $\mu$m plotted in the geometrical parametrisation proposed in
\cite{Lacasa2012a}. The values of the bispectrum are color-coded from
violet to red. In this parametrisation, each subplot shows a slice of
the bispectrum at a given perimeter, indicated in the bottom left
corner. All triangles within the perimeter bin are shown in the
subplot, with squeezed triangles in the upper left corner, equilateral
triangle in the upper right corner and flat isosceles triangle in the
bottom corner. The geometrical parametrisation permits us to
visualise, at the same time, both the scale and configuration
dependence of the bispectrum. We do not display the 1-galaxy
shot-noise term since it is constant.

\begin{figure}
\centering
\includegraphics[width=\linewidth]{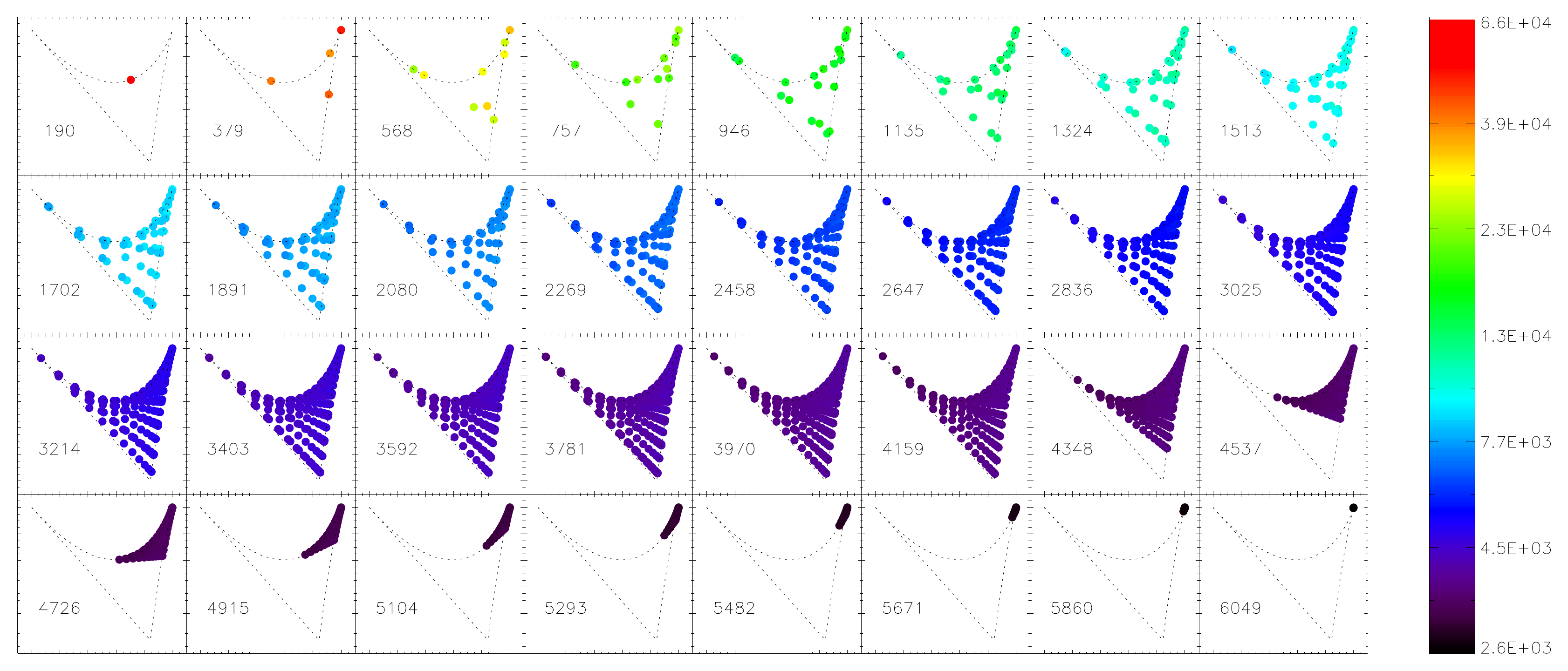}
\caption{The 1-halo term of the CIB bispectrum in the geometrical parametrisation at
  350 $\mu$m = 857 GHz. It has strong scale dependence but few
  dependence on configuration.}
\label{Fig:bl1h_inparam}
\end{figure}

\begin{figure}
\centering
\includegraphics[width=\linewidth]{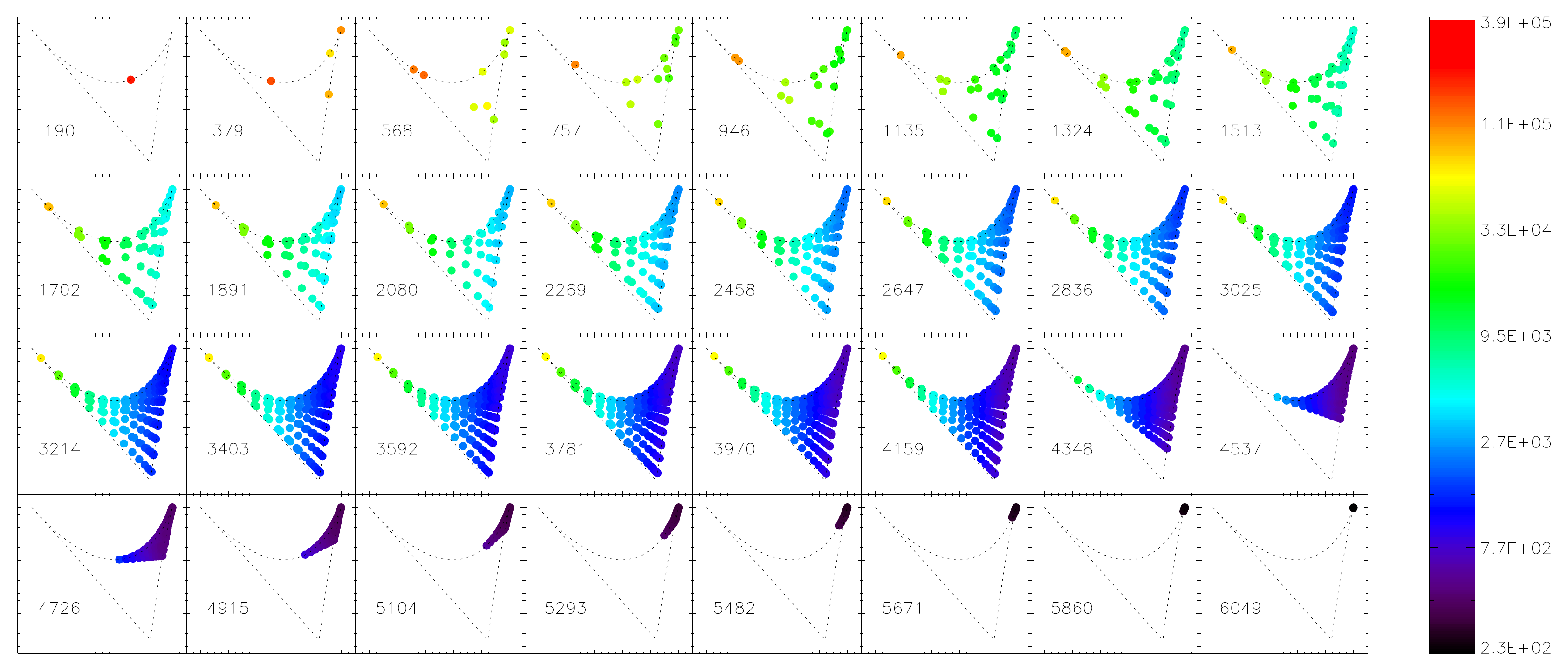}
\caption{The 2-halo term of the CIB bispectrum in the geometrical parametrisation at
  350 $\mu$m = 857 GHz. It exhibits a peak in the squeezed
  configurations (upper left corner of each subplot).}
\label{Fig:bl2h_inparam}
\end{figure}

\begin{figure}
\centering
\includegraphics[width=\linewidth]{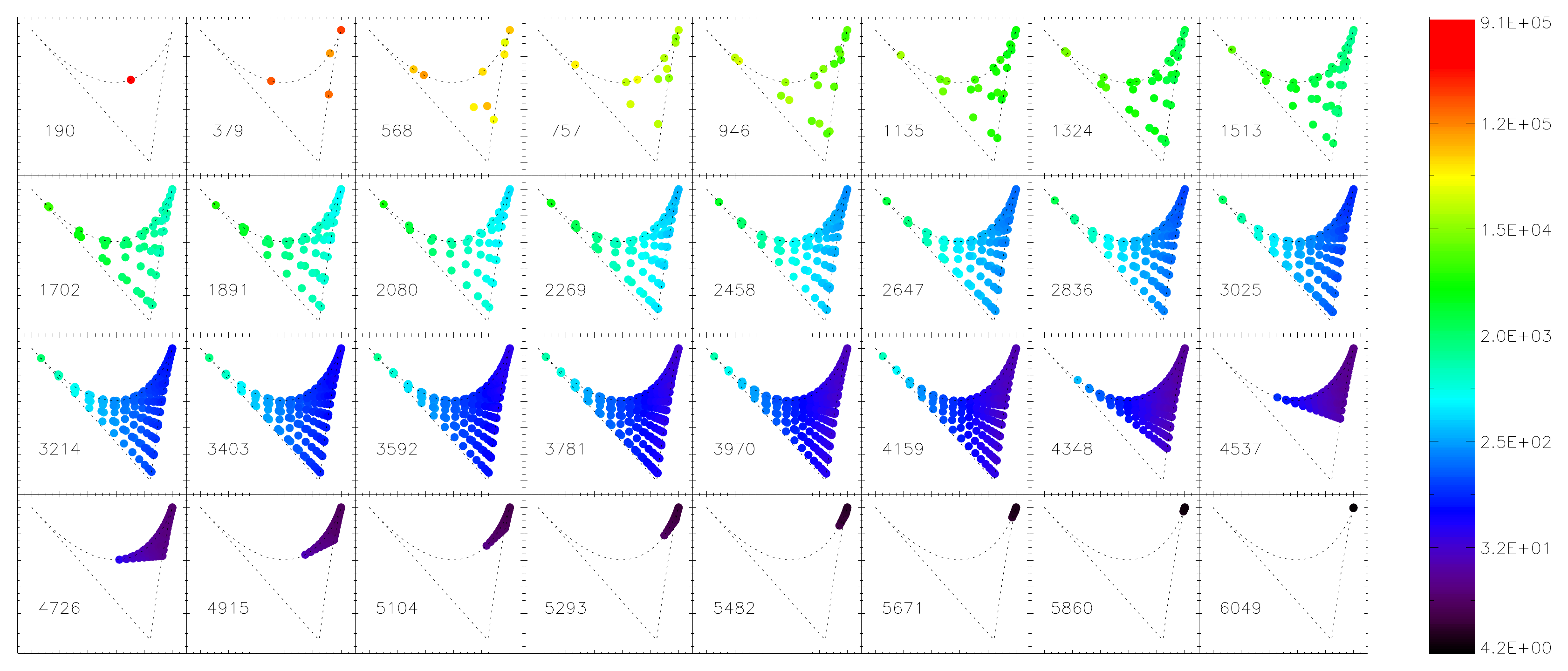}
\caption{The 3-halo term of the CIB bispectrum in the geometrical parametrisation
  at 350 $\mu$m = 857 GHz. It exhibits a peak in the squeezed
  configurations (upper left corner of each subplot).}
\label{Fig:bl3h_inparam}
\end{figure}

\begin{figure}
\centering
\includegraphics[width=\linewidth]{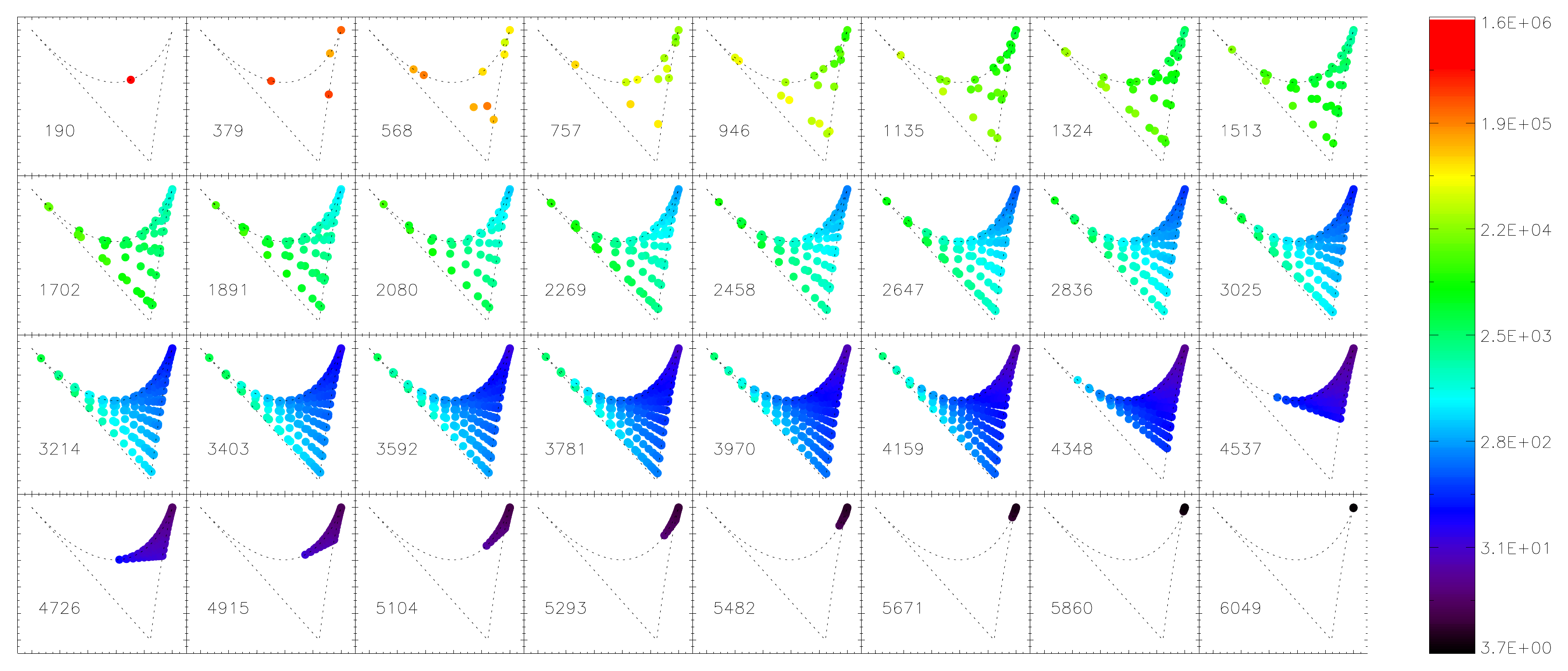}
\caption{The 3-halo cos term of the CIB bispectrum in the geometrical parametrisation
  at 350 $\mu$m = 857 GHz. It exhibits a peak in the squeezed
  configurations (upper left corner of each subplot), and in the flat
  configurations (down corner of each subplot) for some perimeter
  bins.}
\label{Fig:bl3hcos_inparam}
\end{figure}

\begin{figure}
\centering
\includegraphics[width=\linewidth]{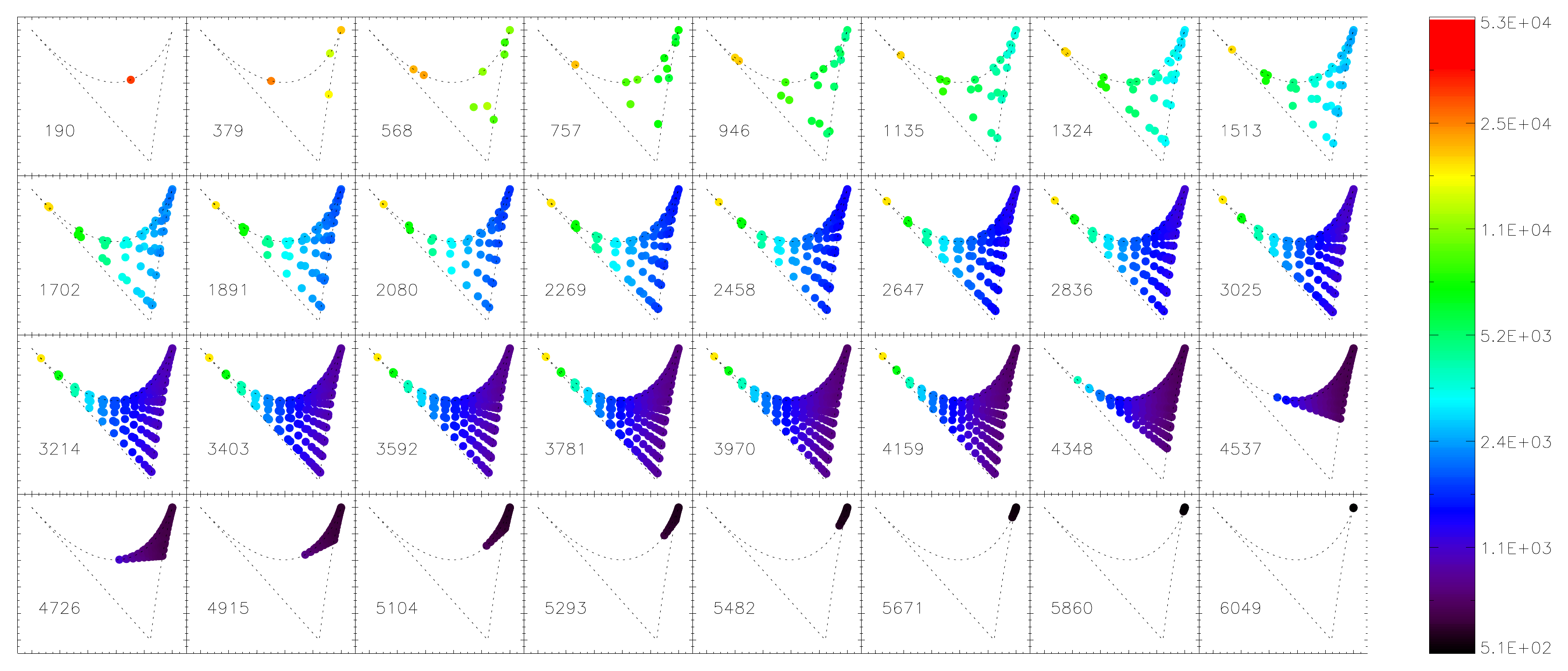}
\caption{2-galaxy shot-noise term of the CIB bispectrum in the
  geometrical parametrisation at 350 $\mu$m = 857 GHz. It exhibits a peak in the
  squeezed configurations (upper left corner of each subplot).}
\label{Fig:blshot2g_inparam}
\end{figure}

As seen already in Fig.\ref{Fig:bl_speconf}, the CIB bispectrum
decreases generally with scale, as it is the case for the power
spectrum. A distinctive feature of the CIB bispectrum is that it peaks
in the squeezed configuration, as already noted by \cite{Lacasa2012a}.
Figures \ref{Fig:bl1h_inparam}--\ref{Fig:blshot2g_inparam} show the
different terms of the CIB bispectrum plotted in the geometrical
parametrisation with the color range adapted to each term to highlight
its variations.

It is worth noting that most terms peak strongly in the squeezed
configurations, except the 1-halo term. The latter has little
configuration dependence but strong scale dependence and is slightly
weaker in squeezed. The 3hcos term from perturbation theory is,
unlike other terms, more important in the flat configuration than in
the equilateral. This is due to the $F^s$ kernel (Eq.\ref{Eq:Fskernel}) which is more important in flat than in
equilateral configurations.


\subsection{Dependencies on the HOD parameters}

\begin{figure*}
\centering 
\includegraphics[width=\linewidth]{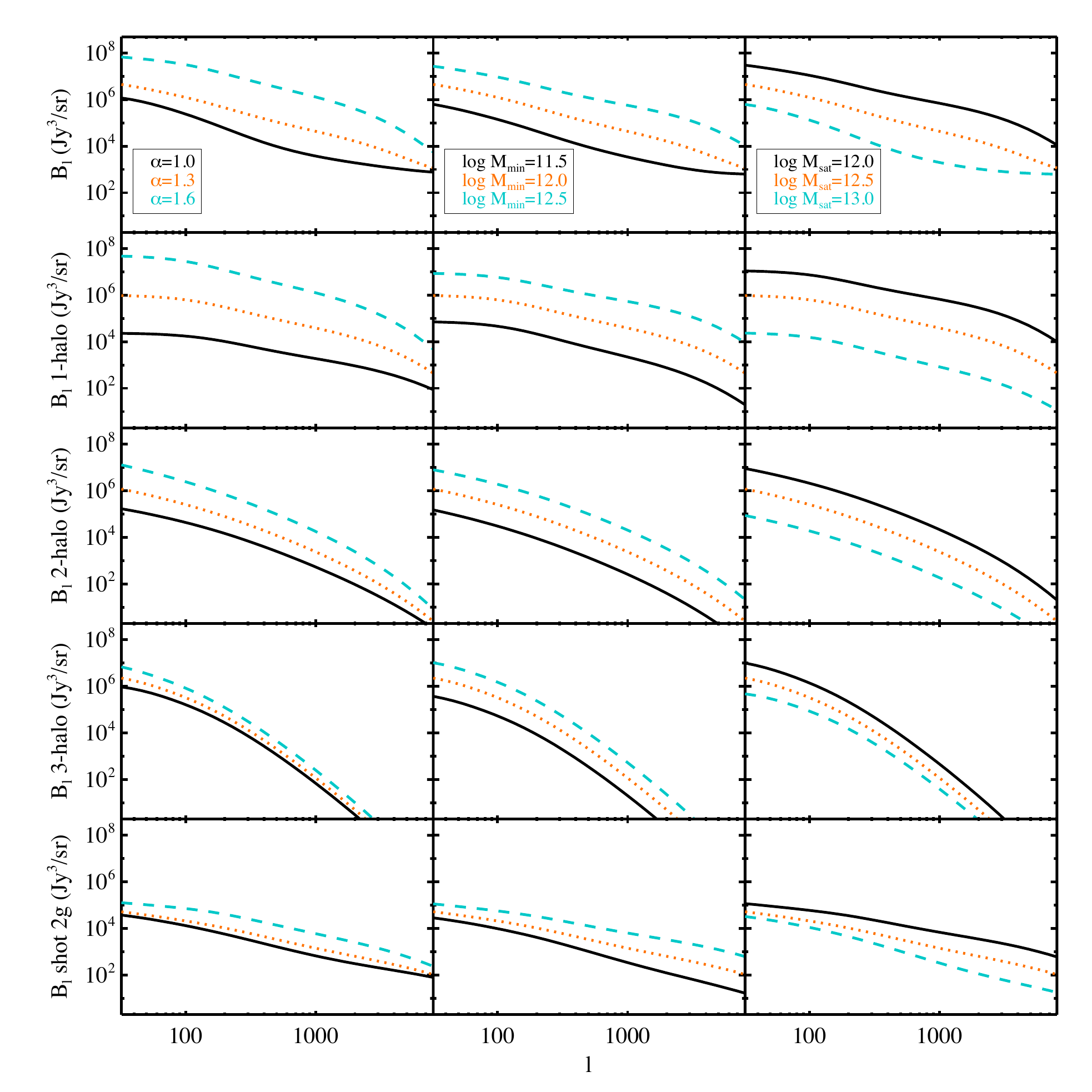}
\caption{Equilateral bispectra at 350 \um~for several sets of the HOD
  parameters. Only one parameter is varied while the others keep their
  fiducial values.} 
\label{fig:bl_vs_hod_param_equi} 
\end{figure*} 

\begin{figure*}
\centering 
\includegraphics[width=\linewidth]{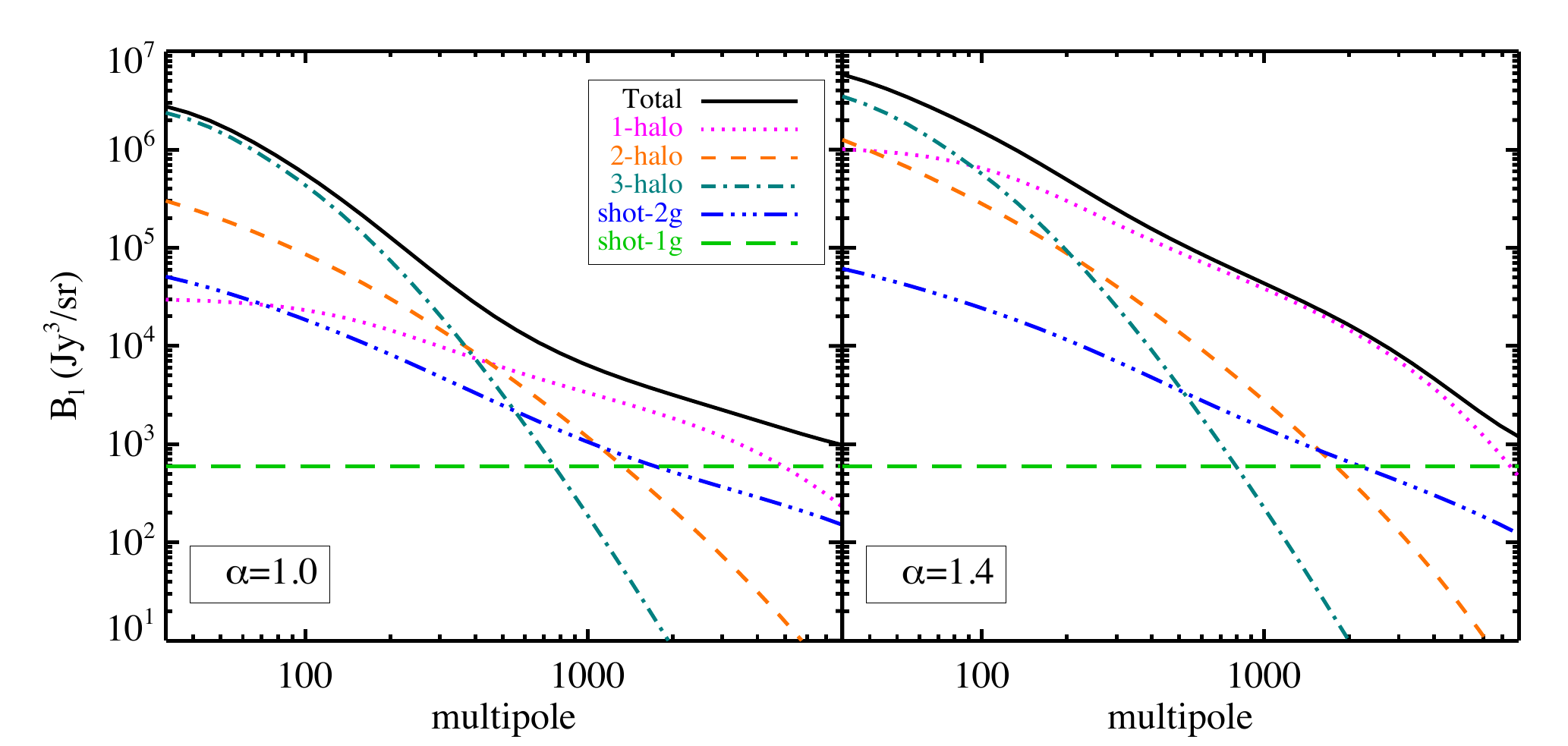} 
\caption{Equilateral bispectrum at 350 \um~with \alphas=1 and \alphas=1.4. The other HOD parameters are fixed at their fiducial values.} 
\label{fig:compare_bl_comp_alphaequi} 
\end{figure*}

We now investigate how the bispectrum and its different components
vary with respect to the HOD parameters. We use a fiducial set of HOD
parameters inspired from previous studies of CIB anisotropies, namely
\citep{Viero2009,Amblard2011,Planck-early-CIB}, and we vary them individually. Only
one parameter is varied, by typically $2 \sigma$ \citep[for
  instance,][]{Planck-early-CIB}, while the others keep their fiducial
values. We consider \alphas=1.3, $\log $ \Mmin = 12, \Msat = 10 \Mmin~and
\sigmalogm = 0.65. As a reminder, increasing (decreasing)
\alphas~leads to a higher (lower) number of satellite galaxies. The
value of \Mmin~rules the mass at which a halo contains a central
galaxy and \Msat~is the average mass of a halo hosting satellite
galaxies.\\

Figure \ref{fig:bl_vs_hod_param_equi} displays the equilateral
bispectrum at 350 \um~and its components. We do not display the
1-galaxy shot noise term as it is independent of the HOD parameters.
First, we see that the amplitude of each component of the bispectrum
is very sensitive to the value of each HOD parameter. Indeed, the
amplitude can increase/decrease by up to two orders of magnitude. For
instance, a higher \alphas~leads to more power at all scales as it
means more satellite galaxies as compared to a lower \alphas. The
shape of the bispectrum terms only varies slightly with the HOD
parameters.  Furthermore, variations of different parameters induce
similar changes on the equilateral bispectrum suggesting a
degeneracies between the HOD parameters. The degeneracy and how it
could be broken are discussed in Paper2.  \\
It is interesting to notice that \alphas~induces strong
variations in the relative contributions of each term of the
bispectrum. Fig. \ref{fig:compare_bl_comp_alphaequi} shows each
component of the bispectrum for two values of \alphas. For
\alphas=1.4 the 1-halo term dominates all the other contributions on
nearly all angular scales.  The bispectrum thus
appears much more sensitive to \alphas~than the power spectrum. This
effect might help to alleviate the tension that exists today about
the measured values of \alphas~(see Paper2).


\subsection{Halos contribution}

\begin{figure*}
\begin{center}
\includegraphics[width=\linewidth]{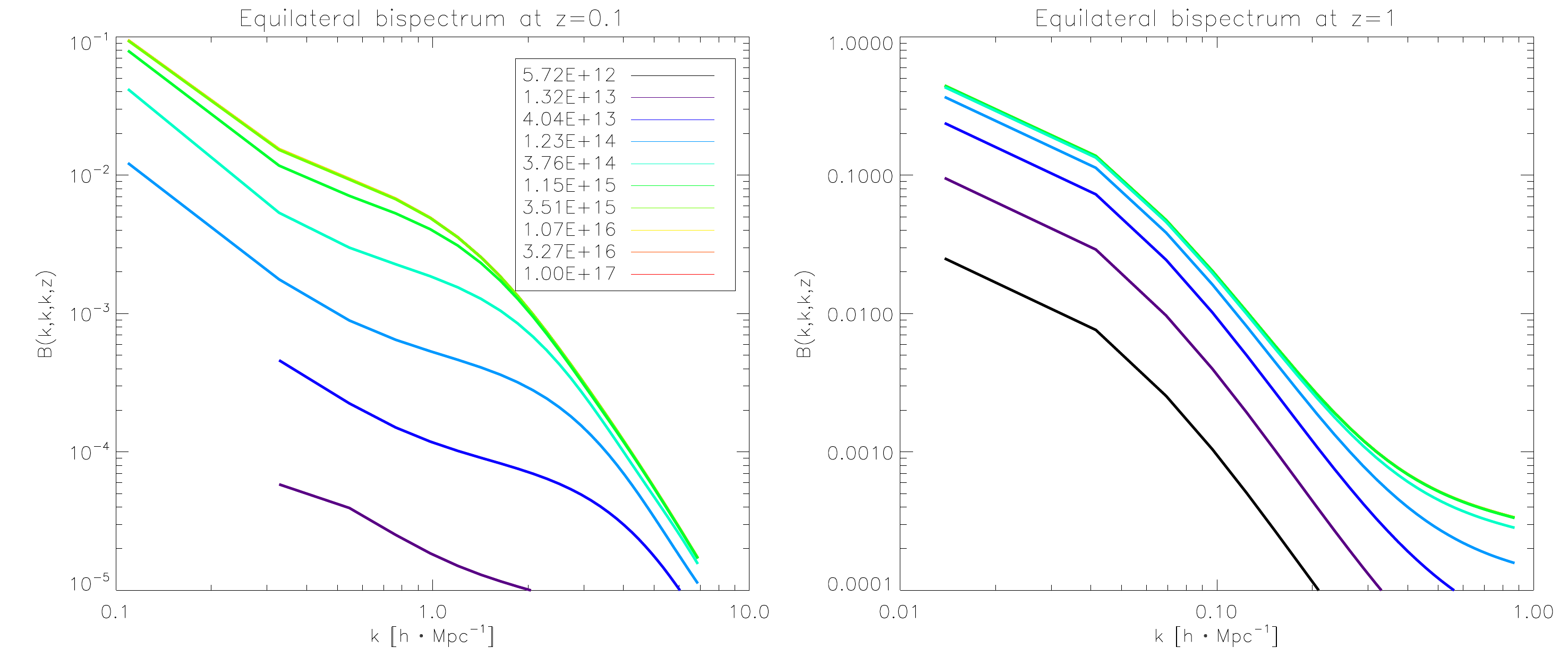}
\caption{Total galaxy equilateral bispectrum at redshift 0.1 and 1, as a function of the halo
  mass upper cut-off from $M_\mathrm{cut} = 5.7 \, 10^{12}\Msun$ to $10^{17}\Msun$.}
\label{Fig:mass-contrib_Bk}
\end{center}
\end{figure*}

We can now focus on halo contribution to the 3D galaxy bispectrum. Most
terms, except the 1-halo term, are not a simple integration over halo
mass, but include cross-terms between halos of different
masses. The 3D galaxy bispectrum cannot be simply divided into a sum of
contributions of different mass bins. We may focus on the
dependence of the bispectrum on the halo-mass upper cut-off.
In Fig.\ref{Fig:mass-contrib_Bk}, we display the total 3D galaxy
bispectrum in the equilateral configuration, as a function of
the halo-mass upper cut-off, respectively at $z$=0.1 and $z$=1.

At $z$=1, we see that the equilateral bispectrum saturates for a
cut-off at a few $10^{14} \, \mathrm{M}_\odot$, except at small scales
where saturation is reached for $\sim 10^{15} \, \mathrm{M}_\odot$. So
at this redshift, halos with masses larger than a few $10^{15} \,
\mathrm{M}_\odot$ contribute negligibly to the bispectrum, as there
are too few of them. At $z$=0.1, massive halos contribute more importantly to
the bispectrum, which saturates at a cut-off $\sim 10^{15} \,
\mathrm{M}_\odot$ on small scales and at $\sim 4 \!\times\!10^{15} \,
\mathrm{M}_\odot$ on large scales. This reflects also the increase in
number of massive halos at low redshift.\\

We checked each of the bispectrum terms and found that the 1-halo term
is the most sensitive to massive halos, while the 3-halo terms (both
3h and 3hcos) are the least sensitive, saturating at a few $10^{14} \,
\mathrm{M}_\odot$ even at low redshift. This behaviour is expected
since the 3-halo terms involve the product of three halo mass
functions. This penalises massive halos in the tail of the mass
function as there are few of them. On the contrary, the 1-halo term
involves only one mass function, and it gives more weight to massive
halos containing more galaxies (see $N_\mathrm{gal}(M)^3$ weight).

The mass contribution to the angular bispectrum depends on the galaxy
emissivities which give weights to each redshift. It hence depends on
the specific galaxy evolution model chosen. This is discussed in
details in the companion article \cite[paper2]{Penin2013}

\section{Discussion}\label{Sect:discussion}

\subsection{Comparison with empirical prescription} 
A simple empirical prescription for the CIB bispectrum based on its
power spectrum was proposed by \cite{Lacasa2012a}. It reads : 
\be
b_{123}^\mathrm{CIB} = \alpha \, \sqrt{C_{\lu}^\mathrm{CIB} \,
  C_{\ld}^\mathrm{CIB} \, C_{\lt}^\mathrm{CIB}} 
\ee 
where $\alpha$ is
a proportionality constant that can be computed with the number
counts of IR galaxies and their flux cut.

\begin{figure}
\begin{center}
\includegraphics[width=\linewidth]{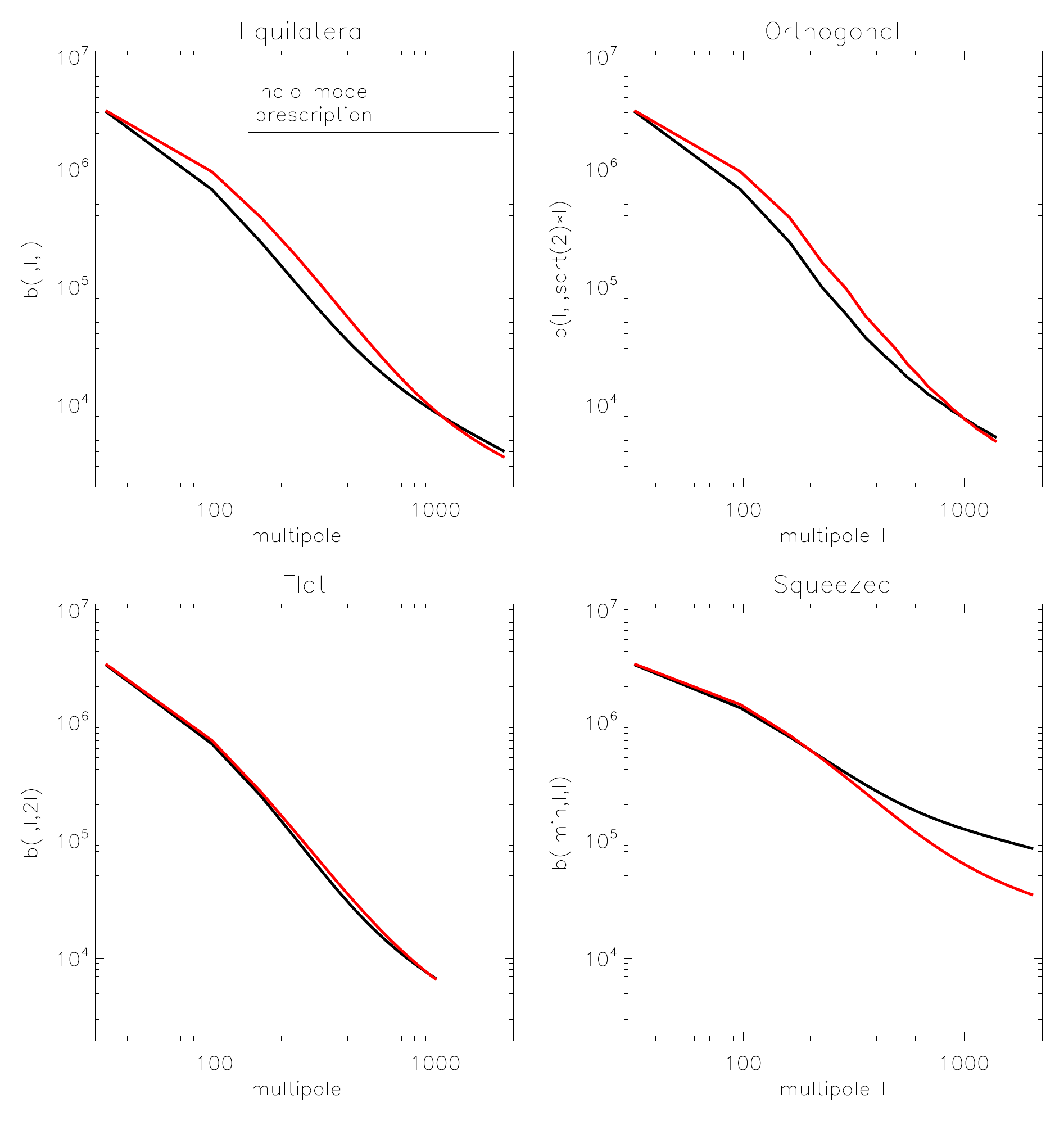}
\caption{CIB bispectrum computed with the halo model (black line) and
  with the prescription (red line) at 350 $\mu$m}
\label{Fig:compwpresp_speconf}
\end{center}
\end{figure}

We compare this prescription with the CIB bispectrum obtained from the
halo model theory. For the prescription, we used the CIB power
spectrum predicted by the halo model with the same parameters as for
the bispectrum, and the best-fit value $\alpha = 2.25 \times 10^{-3}$ (compared to $\alpha = 3 \times 10^{-3}$ found by \citealt{Lacasa2012a} on simulations by \citealt{Sehgal2010}).
 Figure \ref{Fig:compwpresp_speconf} shows both
bispectra at 350 $\mu$m (the prescription is in red and the halo model
in black).

We see that the prescription reproduces reasonably well the shape of
the bispectrum in equilateral, orthogonal and flat
configurations. 
However, the empirical prescription shows an excess of power at intermediate
multipoles $\ell \in [100,1000]$ in the equilateral and isosceles
orthogonal configurations. Finally, the prescription does not recover
the bispectrum in the squeezed configuration, departing from the halo model at $\ell \sim 300$, i.e. when the 2-halo term begins to dominate the squeezed bispectrum (see Fig.\ref{Fig:bl_speconf}).

The empirical prescription therefore gives a reasonable overall fit of the CIB bispectrum, for the considered galaxy emission model and HOD parameters. Furthermore, the prescription gives a separable template (i.e., $b_{123} = f(\ell_1) f(\ell_2) f(\ell_3)$ for some function $f$)\,; it thus provides a convenient way to assess quickly the overall level of CIB non-gaussianity present in a CMB map. This is useful in particular to assess the level of contamination of $f_\mathrm{NL}$ estimation \citep[see][]{Lacasa2012b}. Nevertheless, the empirical prescription does not reproduce completely the theoretical bispectrum derived from the halo model. Additionally in the companion article (Paper2), we show that galaxy evolution models which are indistinguishable at the power spectrum level can produce distinguishable theoretical bispectra with the halo model. On the contrary, for different galaxy models the empirical prescription would give indistinguishable bispectra, as it is based on the power spectra. A full computation of the bispectrum using the halo model is thus necessary if one were to interpret a CIB non-Gaussianity measurement.

\subsection{Comparison with radio sources and CMB bispectra}
At microwave frequencies, several extragalactic signals other than the
CIB are present. In particular, the CMB and radio sources which emit
mostly at low frequencies, $\nu \leq$ 217~GHz \citep{Planck-inter7}.
We thus compare the bispectra of those extragalactic signals with the
CIB theoretical bispectrum derived from the halo model.

Radio sources can be considered distributed randomly on the sky
\citep{Toffolatti1998}, at least for the brightest sources. Hence, the
extragalactic radio background has a white-noise distribution entirely
characterised by the number counts $\frac{\dd N}{\dd S}$. The expected
power spectrum and bispectrum is~:
\bea
C_\ell^\mathrm{RAD} &=& \int_0^{S_\mathrm{cut}} S^2 \, \frac{\dd N}{\dd S} \, \dd S \\
\bl^\mathrm{RAD} &=& \int_0^{S_\mathrm{cut}} S^3 \, \frac{\dd N}{\dd S} \, \dd S
\eea
in Jy$^2$/sr and Jy$^3$/sr$^2$ respectively, with $\frac{\dd N}{\dd
  S}$ in gal/Jy/sr and $S_\mathrm{cut}$ the flux cut. The radio
bispectrum is hence flat, with neither scale nor geometrical
dependence.\\ 
In the following, we use number counts from \cite{Tucci2011}, and flux cuts from \cite{Planck-ERCSC}.

In the standard paradigm, inflation generates close to Gaussian
perturbations which evolve to Gaussian-distributed temperature
fluctuations of the CMB. In the last decade, interest has
increased for the search of CMB non-Gaussianity
\citep[e.g.,][]{Komatsu2011}, as it would be a signature of non-standard
inflation \cite[violating any of the following assumption : slow-roll
single-field inflation with standard kinetic term and Bunch-Davies
initial condition, see ][for a review]{Bartolo2004}, or any
physical process generating the primordial perturbations, or of
non-linear evolution \citep[e.g.,][]{Pitrou2010}. With the recent Planck
measurements, the CMB appears to be very close to a Gaussian field \citep{planck2013-NG}.

Among the many primordial non-Gaussianity shapes, the most studied 
is the `local' non-Gaussianity factor $\fnl$, where the Bardeen potential 
takes the form \citep{Komatsu2001}~:
\be\label{Eq:defnl}
\Phi(\xx) = \Phi_G(\xx) + \fnl\cdot(\Phi_G(\xx)^2 - \langle \Phi_G(\xx)^2 \rangle)
\ee
with $\Phi_G$ the Gaussian part of the potential. This primordial 
non-Gaussianity generates a CMB bispectrum of the form \citep{Komatsu2005} :
\be
\bl^\mathrm{CMB} = \fnl \int r^2 \dd r \; \alpha_{\lu}(r) \, \beta_{\ld}(r) \, \beta_{\lt}(r) + \mathrm{perm.}
\ee
with an integral along the line-of-sight, and filters :
\bea
\alpha_\ell(r) & = & \frac{2}{\pi} \int k^2 \mathrm{d}k \,
g_{T,\ell}(k) \, j_\ell(kr)  \\
\beta_\ell(r) & = & \frac{2}{\pi} \int k^2 \mathrm{d}k \, P(k) \, g_{T,\ell}(k)
\, j_\ell(kr) 
\eea
where $g_{T,\ell}$ is the radiation transfer function, which can 
be computed with a Boltzmann code\footnote{We used CAMB \citep{Lewis2000}}, and $P(k)\propto k^{n_s-4}$ is the primordial power spectrum, with a spectral index $n_s$.\\
The CMB physics, e.g. acoustic peaks and Silk damping, is encoded into 
its bispectrum thanks to the radiation transfer function.

\begin{figure}
\begin{center}
\includegraphics[width=\linewidth]{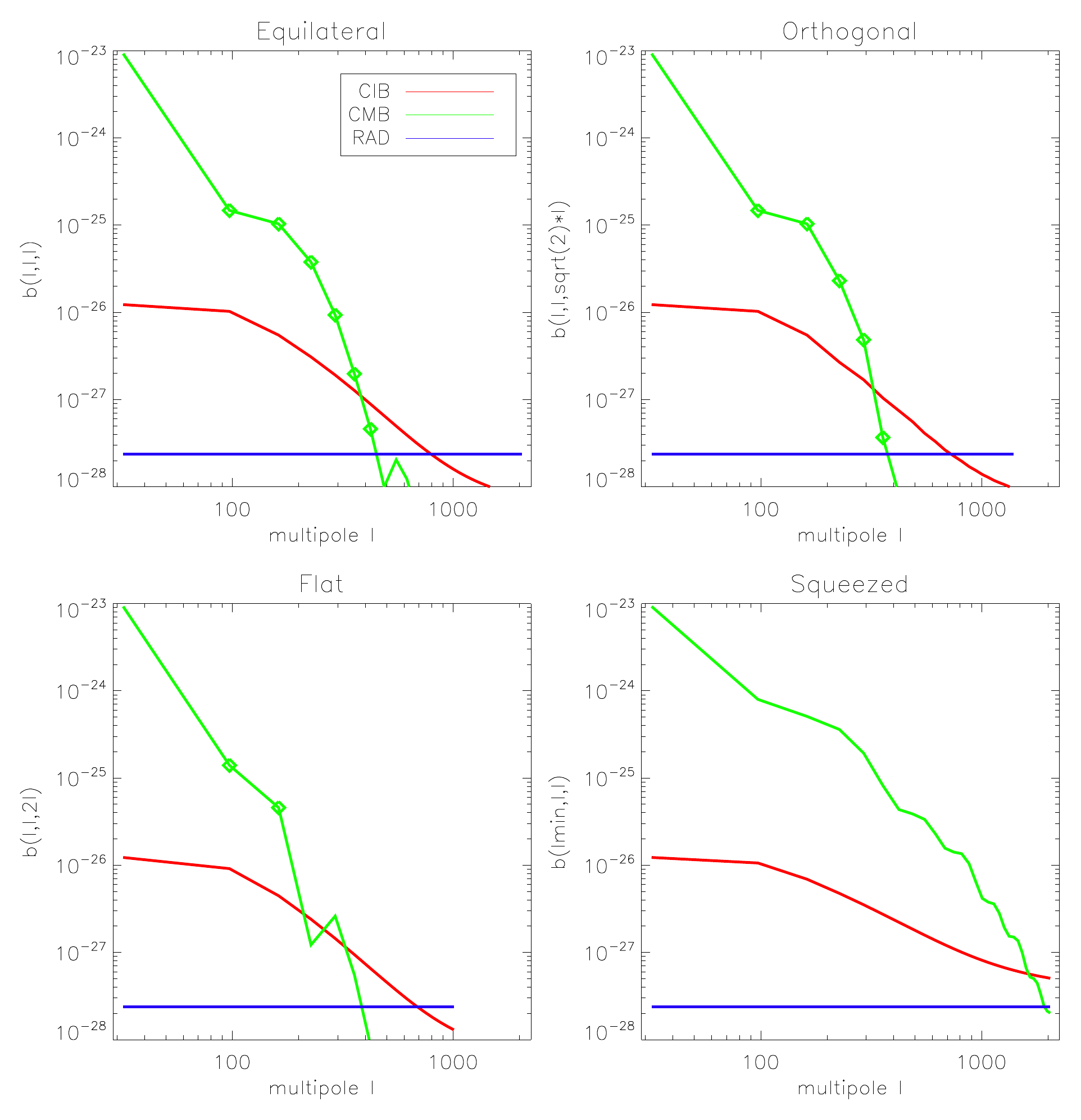}
\caption{CIB (red line), radio (blue line) and CMB (green line)
  bispectra at 220 GHz in dimensionless units $\Delta T/T$. The latter
  is for $f_\mathrm{NL}=1$, and we plot its absolute value as it is
  mostly negative, diamond indicate positive points.}
\label{Fig:speconf_radircmb}
\end{center}
\end{figure}

We show, in Fig.\ref{Fig:speconf_radircmb}, the bispectra of radio
sources and the CMB for $f_\mathrm{NL}=1$ together with the CIB
bispectrum derived from the halo model, in units of relative
temperature elevation $\Delta T/T$, at 1380 $\mu$m = 220 GHz.
At this frequency, radio and infrared point-sources have
comparable contributions, whereas radio sources dominate at lower
frequencies and infrared sources dominate at higher frequencies.

We see that the CMB dominates on large angular scales, but plummets at
high multipoles and becomes negligible for $\ell \geq 500$, except in
the squeezed configuration where the CMB dominates at all
scales. Indeed for local type non-Gaussianity, the CMB bispectrum
peaks strongly in the squeezed limit \citep{Bucher2010}. The CIB
bispectrum also peaks on large angular scales, albeit less strongly
than the CMB. It dominates the radio bispectrum up to $\ell\sim
700-800$ and becomes negligible afterwards except in the squeezed
limit where it dominates the radio over the whole multipole range.
Based on this simple comparison, it seems that detecting the CIB
bispectrum can be possible above 220 GHz.  At 220 GHz, if most of the CMB can be removed by a component separation method
\citep[which estimates a CMB map through multifrequency analysis, see][for a review]{Delabrouille2007}, the detection of the CIB bispectrum is possible at $\ell \leq 700$. Furthermore, the application of a lower flux cut would lower the level of the radio bispectrum and uncover the CIB one. For instance, the CIB bispectrum has been detected in the South Pole Telescope data in a multifrequency analysis using 95, 150 and 220 GHz channels \citep{Crawford2013}, and in the Planck data with a signal-to-noise ratio S/N=5.8 at 217 GHz \citep{Planck2013-CIB}.


\section{Conclusion}\label{Sect:conclusion}

We have presented a framework which allows to predict galaxy
clustering at all orders with the halo model including shot-noise consistently. 
We have developed a new diagrammatic method which allows a clear representation and understanding of the different terms involved in the computation of the 3D galaxy polyspectrum. It further allows to avoid cumbersome computation at high orders by replacing them with diagram drawings. 
This diagrammatic framework is adaptable to different
galaxy-tracing signals and we apply it to the CIB. The latter being integrated over a large range of redshift, we show how the polyspectrum of the CIB anisotropies is projected on the celestial sphere. We further show how to account for the particular case of the shot-noise terms.\\
This framework allows us to compute the CIB angular bispectra at any frequency. We have investigated how the different terms of the resulting CIB bispectrum depend on the scale and on the triangle configuration. We recover that the total bispectrum peaks in the squeezed limit as it is also the case for primordial non-Gaussianity of the local-type. We discuss how the terms of the CIB bispectrum vary with the halo occupation distribution parameters. We show that they vary similarly with respect to the different HOD parameters indicating degeneracies. Furthermore, we show that the bispectrum is much more sensitive to the variation of these parameters than the power spectrum.\\
We explore the halo mass contribution to each term of the 3D galaxy bispectrum, recovering that the 1-halo term gives more weight to massive halos compared to the 2- and 3-halo terms. The halo mass contribution of the angular CIB bispectrum depends on the specific galaxy evolution model which is examined in Paper2.\\
Our predictions are finally compared to a previously proposed empirical prescription and to the bispectrum of radio galaxies and that of the CMB assuming a local-type primordial NG. First, we find an overall agreement with the prescription, although the halo model is needed for an accurate description of the bispectrum, in particular, in the squeezed configuration. Second, we show that the detection of the CIB bispectrum is possible at frequencies above 220 GHz, where the CIB bispectrum is contaminated\,; this detection has indeed been performed by SPT and Planck recent results. \\
This physically-based model opens up the possibility to use, in the future, information present in NG measurement to constrain CIB models so as to extract a maximum of information of present and future surveys.


\section*{Acknowledgments}
The authors wish to acknowledge O. Dor\'{e}, G. P. Holder, G. Lagache, 
C. Porciani, S. Prunet, C. Schimd, E. Sefusatti, J. Tinker, and
A. Wetzel for useful discussions. We acknowledge M. B\'ethermin
for computation of the redshift cut-off discussed in
Sect.\ref{Sect:reglowz} and M. Langer for a thorough reading of the manuscript.


\appendix

\section{Derivation of the galaxy power spectrum equations}\label{App:derivgalpowsp}
Using Eq.\ref{Eq:ngalx}, the computation of the 2-point correlation function of $\delta_\mathrm{gal}$
\be
\zeta_\mathrm{2pt}(\xx_1-\xx_2) = \frac{\langle n_\mathrm{gal}(\xx_1) \, n_\mathrm{gal}(\xx_2)\rangle - \overline{n}_\mathrm{gal}^2}{\overline{n}_\mathrm{gal}^2} = \langle \delta_\mathrm{gal}(\xx_1) \, \delta_\mathrm{gal}(\xx_2) \rangle
\ee
yields a double sum over halo and galaxy indexes $\displaystyle\sum_{i_1,i_2} \sum_{j_1,j_2}$ which can be split in three terms~: $\displaystyle\sum_{i_1 \neq i_2} \sum_{j_1\neq j_2}$ (2-halo term), $\displaystyle\sum_{i_1=i_2} \sum_{j_1\neq j_2}$ (1-halo 2-galaxy term) and $\displaystyle\sum_{i_1 = i_2} \sum_{j_1 = j_2}$ (1-halo 1-galaxy term). So that :
\be
\zeta_\mathrm{2pt}(\xx_1-\xx_2) = \zeta_\mathrm{2pt}^\mr{2h}(\xx_1-\xx_2) + \zeta_\mathrm{2pt}^\mr{1h-2g}(\xx_1-\xx_2) +\zeta_\mathrm{2pt}^\mr{1h-1g}(\xx_1-\xx_2)
\ee
with computations giving :
\ba
\nonumber\zeta_\mathrm{2pt}^\mr{2h}(\xx_1-\xx_2) &= \int\! \dd M_{1} \, \frac{\langle N_\mathrm{gal}(M_{1})\rangle \,\left.\frac{\dd n_\mathrm{h}}{\dd M}\right|_{M_1}}{\overline{n}_\mathrm{gal}} \\
\nonumber & \ \times \int\! \dd M_{2} \, \frac{\langle N_\mathrm{gal}(M_{2})\rangle\, \left.\frac{\dd n_\mathrm{h}}{\dd M}\right|_{M_2}}{\overline{n}_\mathrm{gal}}\\
\nonumber&\ \times \!\int\! \dd^3\xx'_{12} \, u(\xx_{1}-\xx'_{1} | M_{1}) \, u(\xx_{2}-\xx'_{2} | M_{2}) \\
\label{Eq:2ptcf2h}& \ \times \zeta^\mathrm{halo}_\mathrm{2pt} (\xx'_1-\xx'_2 | M_1 , M_2) \\
\nonumber \zeta_\mathrm{2pt}^\mr{1h-2g}(\xx_1-\xx_2) &= \!\int\! \dd M \, \frac{\dd n_\mathrm{h}}{\dd M} \frac{\langle N_\mathrm{gal}(N_\mathrm{gal}-1)\rangle}{\overline{n}_\mathrm{gal}^2}\\
\label{Eq:2ptcf1h2g} & \ \times \int \dd^3\xx_\mathrm{h} \, u(\xx_1-\xx_\mathrm{h} | M) \, u(\xx_2-\xx_\mathrm{h} | M)\qquad\\
\label{Eq:2ptcf1h1g}\zeta_\mathrm{2pt}^\mr{1h-1g}(\xx_1-\xx_2) &= \frac{\delta^3(\xx_1-\xx_2)}{\overline{n}_\mathrm{gal}}
\ea
where $\frac{\dd n_\mathrm{h}}{\dd M}$ is the number of halos with mass M per comoving volume a.k.a. the halo mass function, $u(\xx | M)$ is the halo profile (with integral normalised to unity), and $\zeta^\mathrm{halo}_\mathrm{2pt} (\xx | M_1 , M_2)$ is the halo correlation function conditioned to masses $M_1$ and $M_2$. In this article we use the Sheth \& Tormen mass function \citep{ShandT99} and the associated bias functions, as it is the most recent one for which the second order bias is available.\\
At tree-level, the halo correlation function takes the form \citep[see][and Appendix \ref{App:derivhalo3ptcf}]{Cooray-Sheth2002}~:
\be
\zeta^\mathrm{halo}_\mathrm{2pt} (\xx | M_1 , M_2) = b_1(M_1) \, b_1(M_2) \,\zeta^\mathrm{lin}_\mathrm{2pt} (\xx)
\ee
where $b_1(M)$ is the first order halo bias and $\zeta^\mathrm{lin}$ is the dark matter correlation function at linear/first order in perturbation theory.

The correlation functions defined by Eq.\ref{Eq:2ptcf2h}\&\ref{Eq:2ptcf1h2g} involve convolutions in real space, which become multiplications in Fourier space. Hence the galaxy power spectrum --defined by $\zeta_\mathrm{2pt}(\mathbf{x}) = \int \frac{\dd^3\kk}{(2\pi)^3} P(k) \, e^{i\kk\cdot\mathbf{x}}$-- becomes :
\be
P_\mathrm{gal}(k)=P^\mathrm{1h}_\mathrm{gal}(k) + P^\mathrm{2h}_\mathrm{gal}(k) + P^\mathrm{shot}_\mathrm{gal}(k)
\ee
where the shot-noise contribution (corresponding to the 1h1g term) is examined in more detail in Sect.\ref{Sect:shot-noise}.\\
The 1-halo contribution is
\be
P^\mathrm{1h}_\mathrm{gal}(k) = \int\! \dd M \,\frac{\dd n_\mathrm{h}}{\dd M}\frac{\langle N_\mathrm{gal}(M) (N_\mathrm{gal}(M)-1)\rangle}{\overline{n}_\mathrm{gal}^2} \left|u(k|M)\right|^2
\ee
And the 2-halo contribution
\ba
\nonumber P^\mathrm{2h}_\mathrm{gal}(k) &= \!\int\! \dd M_{1} \, \frac{\langle N_\mathrm{gal}(M_{1})\rangle \,\left.\frac{\dd n_\mathrm{h}}{\dd M}\right|_{M_1}}{\overline{n}_\mathrm{gal}}\,u(k|M_1) \\
\nonumber & \ \times \!\int\! \dd M_{2} \, \frac{\langle N_\mathrm{gal}(M_{2})\rangle \, \left.\frac{\dd n_\mathrm{h}}{\dd M}\right|_{M_2}}{\overline{n}_\mathrm{gal}}\,u(k | M_2) \, P_\mathrm{halo}(k | M_1 , M_2) \\
&= P_\mathrm{lin}(k) \left(\int \dd M \, \frac{\langle N_\mathrm{gal}(M)\rangle \, \frac{\dd n_\mathrm{h}}{\dd M}}{\overline{n}_\mathrm{gal}} \, b_1(M) \, u(k|M) \right)^2
\ea
where, as precedently, all redshift dependence are implicit to simplify notations.


\section{Derivation of the galaxy bispectrum equations}\label{App:derivgalbisp}
Similarly to the 2-point correlation function, using Eq.\ref{Eq:ngalx}, the computation of the 3-point correlation function of $\delta_\mathrm{gal}$ can be split in six terms : $\displaystyle\sum_{i_1 \neq i_2 \neq i_3} \sum_{j_1\neq j_2 \neq j_3}$ (3-halo term), 
$\displaystyle\sum_{i_1=i_2 \neq i_3} \sum_{j_1\neq j_2 \neq j_3} +\mathrm{perm.}$ (2-halo 3-galaxy term), $\displaystyle\sum_{i_1=i_2 \neq i_3} \sum_{j_1=j_2 \neq j_3} +\mathrm{perm.}$ (2-halo 2-galaxy term), 
$\displaystyle\sum_{i_1=i_2=i_3} \sum_{j_1\neq j_2 \neq j_3} +\mathrm{perm.}$ (1-halo 3-galaxy term), 
$\displaystyle\sum_{i_1=i_2=i_3} \sum_{j_1=j_2 \neq j_3} +\mathrm{perm.}$ (1-halo 2-galaxy term), 
and finally $\displaystyle\sum_{i_1=i_2=i_3} \sum_{j_1=j_2=j_3} +\mathrm{perm.}$ (1-halo 1-galaxy term). So that:
\ba
\nonumber\zeta_\mathrm{3pt}(\xx_1,\xx_2,\xx_3) &= \zeta_\mathrm{3pt}^\mr{3h}(\xx_1,\xx_2,\xx_3) + \zeta_\mathrm{3pt}^\mr{2h-3g}(\xx_1,\xx_2,\xx_3) \\
\nonumber & +\zeta_\mathrm{3pt}^\mr{2h-2g}(\xx_1,\xx_2,\xx_3)+\zeta_\mathrm{3pt}^\mr{1h-3g}(\xx_1,\xx_2,\xx_3) \\
& + \zeta_\mathrm{3pt}^\mr{1h-2g}(\xx_1,\xx_2,\xx_3) +\zeta_\mathrm{3pt}^\mr{1h-1g}(\xx_1,\xx_2,\xx_3)
\ea
with computations giving :
\ba
\nonumber \zeta_\mathrm{3pt}^\mr{3h}(\xx_1,\xx_2,\xx_3) &=\!\int\! \dd M_{123}\left[\frac{\langle N_\mathrm{gal}(M_i)\rangle \, \left.\frac{\dd n_\mathrm{h}}{\dd M}\right|_{M_i}}{\overline{n}_\mathrm{gal}}\right]_{i=123}\\
\nonumber & \times \int\! \dd^3\mathbf{x}'_{123} \left(u(\mathbf{x}_{i}-\mathbf{x}'_{i} | M_{i})\right)_{i=123}\\
\label{Eq:3ptcf3h} &\quad \times \zeta_\mathrm{3pt}^\mathrm{halo}(\mathbf{x}'_{123} | M_{123})
\ea
\ba
\nonumber\zeta_\mathrm{3pt}^\mr{2h-3g}(\xx_1,\xx_2,\xx_3)&=\int \dd M_1 \frac{\langle N_\mathrm{gal}(M_1) \left(N_\mathrm{gal}(M_1)-1\right)\rangle\, \left.\frac{\dd n_\mathrm{h}}{\dd M}\right|_{M_1}}{\overline{n}_\mathrm{gal}^2}\\
\nonumber & \ \times \int \dd M_3 \frac{\langle N_\mathrm{gal}(M_3)\rangle\, \left.\frac{\dd n_\mathrm{h}}{\dd M}\right|_{M_3}}{\overline{n}_\mathrm{gal}}\\
\nonumber& \ \times\!\!\int\! \dd^3\mathbf{x}'_{13} \,u(\mathbf{x}_{1}-\mathbf{x}'_{1} | M_{1})\;\! u(\mathbf{x}_{2}-\mathbf{x}'_{1} | M_{1})\\
\nonumber& \quad \times u(\mathbf{x}_{3}-\mathbf{x}'_{3} | M_{3}) \,\zeta^\mathrm{halo}_\mathrm{2pt}(\mathbf{x}'_{1}-\mathbf{x}'_{3} | M_1 M_3)\\ 
\label{Eq:3ptcf2h3g} &\qquad+\,\mathrm{perm.}
\ea
\ba
\label{Eq:3ptcf2h2g}\zeta_\mathrm{3pt}^\mr{2h-2g}(\xx_1,\xx_2,\xx_3)&=\frac{\zeta^\mr{2h-2g}_\mathrm{2pt}(\xx_1-\xx_3)\,\delta^{(3)}(\xx_1-\xx_2)}{\overline{n}_\mathrm{gal}}+ \mathrm{perm.}
\ea
\ba
\nonumber \zeta_\mathrm{3pt}^\mr{1h-3g}(\xx_1,\xx_2,\xx_3)&=\int \dd M \, \frac{\langle N_\mathrm{gal}\left(N_\mathrm{gal}-1\right)\left(N_\mathrm{gal}-2\right)\rangle}{\overline{n}_\mathrm{gal}^3} \, \frac{\dd n_\mathrm{h}}{\dd M} \\
\nonumber &\quad \times \int \dd^3\mathbf{x}' \, u(\mathbf{x}_1-\mathbf{x}'|M) u(\mathbf{x}_2-\mathbf{x}'|M) \\
\label{Eq:3ptcf1h3g} & \qquad \times u(\mathbf{x}_3-\mathbf{x}'|M)
\ea
\ba
\label{Eq:3ptcf1h2g} \zeta_\mathrm{3pt}^\mr{1h-2g}(\xx_1,\xx_2,\xx_3)&= \frac{\zeta^\mr{1h-2g}_\mathrm{2pt}(\xx_1-\xx_3)\,\delta^{(3)}(\xx_1-\xx_2)}{\overline{n}_\mathrm{gal}}+ \mathrm{perm.}
\ea
\ba
\label{Eq:3ptcf1h1g} \zeta_\mathrm{3pt}^\mr{1h-1g}(\xx_1,\xx_2,\xx_3)&= \frac{\delta^{(6)}(\xx_1=\xx_2=\xx_3)}{\overline{n}_\mathrm{gal}^2}
\ea
where $\zeta_\mathrm{3pt}^\mathrm{halo}(\mathbf{x}_{123} | M_{123})$ is the halo 3-point correlation function conditioned to masses $M_1$ $M_2$ and $M_3$, and for which we take (see Appendix \ref{App:derivhalo3ptcf}) :
\ba \label{Eq:halo3ptcf}
\nonumber \zeta^\mathrm{halo}_\mathrm{3pt} (\mathbf{x}_{123} | M_{123}) &= b_1(M_1) \, b_1(M_2) \, b_1(M_3) \,\zeta^\mathrm{DM}_\mathrm{3pt} (\mathbf{x}_{123})\\
\nonumber & \ +\, b_1(M_1) \, b_1(M_2) \, b_2(M_3) \, \zeta^\mathrm{lin}_\mathrm{2pt} (\xx_1-\xx_3)\\
& \qquad \times \zeta^\mathrm{lin}_\mathrm{2pt} (\xx_2-\xx_3) + \mathrm{perm.}
\ea
where $b_2$ is the second order halo bias.

The 3D galaxy bispectrum can be computed from Eq.\ref{Eq:3ptcf3h} to \ref{Eq:3ptcf1h3g} through Fourier Transform 
\ba
\nonumber \zeta_\mathrm{3pt}(\mathbf{x}_1,\mathbf{x}_2,\mathbf{x}_3) &=\int\frac{\dd^3 \kk_{123}}{(2\pi)^9} \, B(k_1,k_2,k_3) \, (2\pi)^3\delta(\kk_{1}+\kk_{2}+\kk_{3}) \\
\nonumber & \qquad \times e^{i(\kk_{1}\cdot\mathbf{x}_{1}+\kk_{2}\cdot\mathbf{x}_{2}+\kk_{3}\cdot\mathbf{x}_{3})}
 \ea
giving :
\ba
\nonumber B_\mathrm{gal}(k_1,k_2,k_3) &= B^\mathrm{1h}_\mathrm{gal}(k_1,k_2,k_3)+B^\mathrm{2h}_\mathrm{gal}(k_1,k_2,k_3)\\
\nonumber &+B^\mathrm{3h}_\mathrm{gal}(k_1,k_2,k_3) + B^\mathrm{shot2g}_\mathrm{gal}(k_1,k_2,k_3) \\
& + B^\mathrm{shot1g}_\mathrm{gal}(k_1,k_2,k_3)
\ea
where we have the 1-halo term :
\ba
\nonumber B^\mathrm{1h}_\mathrm{gal}(k_1,k_2,k_3) &= \int \dd M \, \frac{\langle N_\mathrm{gal} (N_\mathrm{gal}-1)(N_\mathrm{gal}-2)\rangle}{\overline{n}_\mathrm{gal}^3} \frac{\dd n_\mathrm{h}}{\dd M} \\
& \qquad \times u(\kk_1 | M) \, u(\kk_2 | M) \, u(\kk_3 | M)
\ea
The 2-halo term :
\ba
\nonumber B^\mathrm{2h}_\mathrm{gal}(k_1,k_2,k_3) &= \!\int \!\dd M_1 \frac{\langle N_\mathrm{gal} (N_\mathrm{gal}-1)\rangle \, \left.\frac{\dd n_\mathrm{h}}{\dd M}\right|_{M_1}}{\overline{n}_\mathrm{gal}^2}\\
\nonumber & \quad \times \int\! \dd M_3 \frac{\langle N_\mathrm{gal}(M_3)\rangle \, \left.\frac{\dd n_\mathrm{h}}{\dd M}\right|_{M_3}}{\overline{n}_\mathrm{gal}}\\
\nonumber & \quad \times u(\kk_1 | M_1) u(\kk_2 | M_1) u(\kk_3 | M_3) \\
& \quad \times P_\mathrm{halo}(k_3 | M_1 , M_3) \ + \mathrm{perm.}\\
\nonumber &= P_\mathrm{lin}(k_3)\!\!\int \!\dd M_1 \frac{\langle N_\mathrm{gal} (N_\mathrm{gal}-1) \rangle\, \left.\frac{\dd n_\mathrm{h}}{\dd M}\right|_{M_1}}{\overline{n}_\mathrm{gal}^2} \\
\nonumber & \quad \times b_1(M_1) \, u(\kk_1 | M_1) u(\kk_2 | M_1)\\
\nonumber & \quad \times \!\int\! \dd M_3 \frac{\langle N_\mathrm{gal}(M_3)\rangle \, \left.\frac{\dd n_\mathrm{h}}{\dd M}\right|_{M_3}}{\overline{n}_\mathrm{gal}} \, b_1(M_3) \, u(\kk_3 | M_3)\\
& \quad +  \mathrm{perm.}
\ea
And the 3-halo term :
\ba
\nonumber B^\mathrm{3h}_\mathrm{gal}(k_1,k_2,k_3) = {} &  \int \dd M_{123}\left(\frac{N_\mathrm{gal}(M_i)\, \left.\frac{\dd n_\mathrm{h}}{\dd M}\right|_{M_i}}{\overline{n}_\mathrm{gal}}\right)_{i=123} \!\! \left|u(\kk_i | M_i)\right|_{i=123}\\
& \times B_\mathrm{halo}(k_1,k_2,k_3 | M_1,M_2,M_3)\\
\nonumber = {} & B_\mathrm{lin}(k_{123}) \int \dd M_{123} \bigg(\frac{N_\mathrm{gal}(M_i)\, \left.\frac{\dd n_\mathrm{h}}{\dd M}\right|_{M_i}}{\overline{n}_\mathrm{gal}} \\
\nonumber & \qquad \times b_1(M_i) \, u(\kk_i | M_i) \bigg)_{i=123} \\
\nonumber &+\int \dd M_{123}\left(\frac{N_\mathrm{gal}(M_i)\, \left.\frac{\dd n_\mathrm{h}}{\dd M}\right|_{M_i}}{\overline{n}_\mathrm{gal}} \left|u(\kk_i | M_i)\right| \right)_{i=123}\\
\nonumber & \quad \times b_1(M_1) \, b_1(M_2) \, b_2(M_3) \, P_\mathrm{lin}(k_1)P_\mathrm{lin}(k_2)\\
& +\ \mathrm{perm.}
\ea
with
\ba\label{Eq:Bdm}
\nonumber B_\mathrm{DM}(k_1,k_2,k_3) &= \underbrace{2 \, f_\mathrm{NL}\left( \left(\frac{k_3 \, k_H}{k_1 \, k_2}\right)^2 P_\mathrm{lin}(k_1) \, P_\mathrm{lin}(k_2) + \mathrm{perm.} \right)}_{\mathrm{primordial \ NG}}\\
&+\underbrace{2 \, F^s(\mathbf{k_1}, \mathbf{k_2}) \, P_\mathrm{lin}(k_1) \, P_\mathrm{lin}(k_2) + \mathrm{perm.}}_\mathrm{gravity \ non-linearity \ at \ 2PT}
\ea
with
\be
k_H^2 = 4\pi G \, \overline{\rho} \, a^2(t)
\ee
which stems from the Poisson equation linking density contrast to Bardeen potential involved in the definition of $\fnl$ (Eq.\ref{Eq:defnl})\\
and
\be
F^s(\mathbf{k_i},\mathbf{k_j})=\frac{5}{7}+\frac{1}{2}\cos(\theta_{ij})(\frac{k_i}{k_j} +\frac{k_j}{k_i})+\frac{2}{7}\cos^2(\theta_{ij})
\ee
which stems from non-linear evolution at second order in perturbation theory \citep{Fry1984,Gil-Marin2012}.\\
At the scales of interest in this article $k_H \ll k_{1,2,3}$ so that we can neglect the primordial NG term (e.g. compared to the 2PT term in Eq.\ref{Eq:Bdm}).

The equations above can be somewhat simplified if we introduce some notations (where we reintroduced the redshift dependence to show where it intervenes) :
\be
\mathcal{F}_1(k,z) = \int \dd M \frac{\langle N_\mathrm{gal}(M)\rangle}{\overline{n}_\mathrm{gal}(z)} \,\frac{\dd n_\mathrm{h}}{\dd M}(M,z) \, b_1(M,z) \, u(k|M,z)
\ee
\be
\mathcal{F}_2(k,z) = \int \dd M \frac{\langle N_\mathrm{gal}(M)\rangle}{\overline{n}_\mathrm{gal}(z)} \,\frac{\dd n_\mathrm{h}}{\dd M}(M,z) \,b_2(M,z) \, u(k|M,z)
\ee
\ba
\nonumber \mathcal{G}_1(k_1,k_2,z) = {}Ê& \int \dd M \frac{\langle N_\mathrm{gal}(N_\mathrm{gal}-1)\rangle}{\overline{n}_\mathrm{gal}(z)^2} \,\frac{\dd n_\mathrm{h}}{\dd M}(M,z) \\
& \times b_1(M,z) \, u(k_1|M,z) \, u(k_2|M,z)
\ea
and the $F^s$ kernel can also be computed through the formula
\be
F^s(\mathbf{k_\alpha},\mathbf{k_\beta}) = \frac{2 k_\gamma^4 -5(k_\alpha^4+k_\beta^4)+3 k_\gamma^2(k_\alpha^2+k_\beta^2) +10 k_\alpha^2k_\beta^2}{28 k_\alpha^2k_\beta^2}
\ee
where $\gamma$ is the third index

With these notations the 2-halo term takes the form :
\ba
\nonumber B^\mathrm{2h}_\mathrm{gal}(k_1,k_2,k_3,z) = {} & \mathcal{G}_1(k_1,k_2,z) \; P_\mathrm{lin}(k_3 , z) \, \mathcal{F}_1(k_3,z) \\
\nonumber & +\ \mathcal{G}_1(k_1,k_3,z)\; P_\mathrm{lin}(k_2 , z) \,\mathcal{F}_1(k_2,z) \\
& +\ \mathcal{G}_1(k_2,k_3,z) \; P_\mathrm{lin}(k_1 ,z ) \, \mathcal{F}_1(k_1,z)
\ea
and the 3-halo term :
\ba\label{Eq:Bk3hwFnot}
\nonumber B^\mathrm{3h}_\mathrm{gal}(k_1,k_2,k_3,z) = {} & \mathcal{F}_1(k_1,z)\, \mathcal{F}_1(k_2,z)\, \mathcal{F}_1(k_3,z)\\
\nonumber & \times \left[F^s(\mathbf{k_1},\mathbf{k_2})\, P_\mathrm{lin}(k_1,z)\, P_\mathrm{lin}(k_2,z) +\mathrm{perm.}\right] \\
\nonumber & +\ \mathcal{F}_1(k_1,z)\, \mathcal{F}_1(k_2,z)\, \mathcal{F}_2(k_3,z)\\
\nonumber & \quad \times P_\mathrm{lin}(k_1 , z)\, P_\mathrm{lin}(k_2 , z) + \mathrm{perm.} \\
\ea
Last, the shot-noise terms are :
\ba
B^\mr{1h-1g}_\mathrm{gal}(k_1,k_2,k_3,z) &= \frac{1}{\overline{n}_\mathrm{gal}^2(z)}\\
B^\mr{1h-2g}_\mathrm{gal}(k_1,k_2,k_3,z) &=\frac{P^\mr{1h-2g}_\mathrm{gal}(k_1)+P^\mr{1h-2g}_\mathrm{gal}(k_2)+P^\mr{1h-2g}_\mathrm{gal}(k_3)}{\overline{n}_\mathrm{gal}(z)}\\
B^\mr{2h-2g}_\mathrm{gal}(k_1,k_2,k_3,z) &=\frac{P^\mr{2h-2g}_\mathrm{gal}(k_1)+P^\mr{2h-2g}_\mathrm{gal}(k_2)+P^\mr{2h-2g}_\mathrm{gal}(k_3)}{\overline{n}_\mathrm{gal}(z)}
\ea


\section{Halo correlation functions}\label{App:derivhalo3ptcf}
We assume that the halo density field follows the local bias scheme \citep{Fry1993} :
\be
\delta_\mathrm{h}(\xx | M) = \sum_{n=1}^{+\infty} \frac{b_n(M)}{n!} \: \delta_\mathrm{DM}(\xx)^n
\ee
where $\delta_\mathrm{DM}$ is the dark  matter density field predicted through perturbation theory, and $b_n(m)$ is the n-th order bias 
\be
b_n(M) = \frac{1}{f(\nu)} \frac{\partial^n f(\nu)}{\partial \delta^n}
\ee
Because of the smallness of $\delta_\mathrm{DM}$, it is sufficient to develop the computation of halo correlation functions to tree-level.\\
Hence the 2-point correlation function conditioned to mass $M_1$ and $M_2$ is :
\ba
\nonumber\zeta^\mr{hh}_\mathrm{2pt}(\xx_1-\xx_2 | M_1, M_2) &\equiv \langle \delta_\mathrm{h}(\xx_1 | M_1) \,  \delta_\mathrm{h}(\xx_2 | M_2) \rangle\\
\nonumber &= b_1(M_1) \, b_1(M_2) \, \langle \delta_\mathrm{DM}(\xx_1) \, \delta_\mathrm{DM}(\xx_2) \rangle\\
\nonumber & \quad +\mathbf{o}\left(\delta^2\right)\\
& \simeq b_1(M_1) \, b_1(M_2) \, \zeta^\mathrm{lin}_\mathrm{2pt}(\xx_1-\xx_2)
\ea
Going to Fourier space gives the power spectrum :
\be
P_{hh}(k) = b_1(M_1) \, b_1(M_2) \, P_\mathrm{lin}(k)
\ee
At tree-level the 3-point correlation function conditioned to mass $M_1$ $M_2$ and $M_3$ is :
\ba
\nonumber\zeta^\mr{hhh}_\mathrm{3pt}(\xx_{123} | M_{123}) = {} & \langle \left(b_1(M_1) \delta_\mathrm{DM}(\xx_1) + \frac{b_2(M_1)}{2!} \delta_\mathrm{DM}(\xx_1)^2\right)\\
 \nonumber & \times \left(b_1(M_2) \delta_\mathrm{DM}(\xx_2)+ \frac{b_2(M_2)}{2!} \delta_\mathrm{DM}(\xx_2)^2\right)\\
\nonumber &  \times \left(b_1(M_3) \delta_\mathrm{DM}(\xx_3) + \frac{b_2(M_3)}{2!} \delta_\mathrm{DM}(\xx_3)^2\right) \rangle\\
\nonumber= {} & b_1(M_1) \, b_1(M_2) \, b_1(M_3) \, \zeta^\mathrm{DM}_\mathrm{3pt}(\xx_1 , \xx_2 , \xx_3) \\
\nonumber&+ \frac{b_2(M_1)}{2!} \, b_1(M_2) \, b_1(M_3) \\
\label{Eq:4ptinzetahhh} & \times \langle \delta_\mathrm{lin}(\xx_1)^2 \, \delta_\mathrm{lin}(\xx_2) \, \delta_\mathrm{lin}(\xx_3)\rangle + 2\;\mathrm{perm.}\\
\nonumber= {} & b_1(M_1) \, b_1(M_2) \, b_1(M_3) \, \zeta^\mathrm{lin}_\mathrm{3pt}(\xx_1 , \xx_2 , \xx_3) \\
\nonumber&+ b_2(M_1) \, b_1(M_2) \, b_1(M_3) \\
\nonumber & \times \bigg[\frac{\langle\delta^2_\mathrm{DM}\rangle}{2!}\,\zeta^\mathrm{lin}_\mathrm{2pt}(\xx_2\!-\!\xx_3) \\
& + \zeta^\mathrm{lin}_\mathrm{2pt}(\xx_1\!-\!\xx_2)\,\zeta^\mathrm{lin}_\mathrm{2pt}(\xx_1\!-\!\xx_3)\bigg] + 2\;\mathrm{perm.}
\ea
where we expanded the 4-point correlation function (at line \ref{Eq:4ptinzetahhh}) through Wick's theorem as the field is close to Gaussian, and where $\zeta^\mathrm{DM}_\mathrm{3pt}(\xx_1 , \xx_2 , \xx_3)$ contains two contributions~: primordial non-Gaussianity which is observationally constrained to be small, and non-Gaussianity generated by the non-linearity of gravity at second-order in perturbation theory.\\
Going to Fourier space gives the bispectrum (for $k_1 , k_2 , k_3 \neq 0$) :
\ba
\nonumber B_\mr{hhh}(k_{123} | M_{123}) = {} &  b_1(M_1) \, b_1(M_2) \, b_1(M_3) \, B_\mathrm{DM}(k_1 , k_2 , k_3) \\
\nonumber & + b_2(M_1) \, b_1(M_2) \, b_1(M_3) \, P_\mathrm{lin}(k_2) \, P_\mathrm{lin}(k_3)\\
& + 2\;\mathrm{perm.}
\ea


\section{Polyspectra in Fourier space and on the sphere}\label{App:polyspectradef}

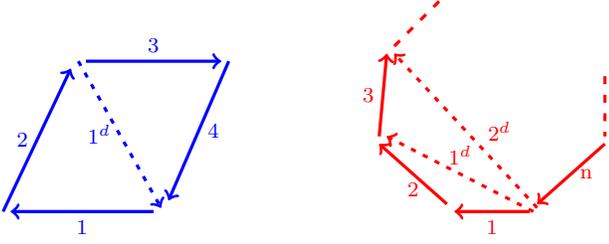
\begin{figure}
\begin{center}
\begin{tikzpicture}
\draw [->, very thick,blue] (0,-1) -- node[left] {2} (0.9,0.9);
\draw [->, very thick, dash pattern=on 2pt off 3pt,blue] (1,1) -- node[left] {$1^d$} (2.1,-0.95); 
\draw [->, very thick,blue] (1.1,1) -- node[above] {3} (2.9,1);
\draw [->, very thick,blue]  (3,1) -- node[right] {4} (2.2,-0.85);
\draw [->, very thick,blue] (2,-1) -- node[below] {1} (0.1,-1);
\draw [very thick,dash pattern=on 3pt off 4pt,red] (8,0.8) --  (8,0);
\draw [->, very thick,red] (8,-0.1) -- node[right] {n} (7.1,-0.9);
\draw [->, very thick,red] (7,-1) -- node[below] {1} (6,-1);
\draw [->, very thick,red] (5.9,-0.9) -- node[left,below] {2} (5,-0.1);
\draw [->, very thick,red] (5,0) -- node[left] {3} (5.1,1.1);
\draw [very thick,dash pattern=on 3pt off 4pt,red] (5.2,1.2) --  (5.8,1.8);
\draw [->, very thick, dash pattern=on 2pt off 3pt,red] (7.05,-1) -- node[above] {$1^d$} (5.1,0);
\draw [->, very thick, dash pattern=on 2pt off 3pt,red] (7.1,-0.95) -- node[right=5pt] {$2^d$} (5.2,1.1);
\end{tikzpicture}
\caption{Parametrisation of the 2D polyspectrum, with diagonals in dashed lines, at \textcolor{blue}{fourth} (trispectrum) and \textcolor{red}{$n$-th} order.}
\label{Fig:polygons2D}
\end{center}
\end{figure}

Let $a_{\kk}$ be the Fourier transform of a random field. For example this may be the 3D galaxy density field, in which case $a_{\kk}=\delta_\mathrm{gal}(\kk , z)$. The polyspectrum of order $n$ is then defined via the connected correlation function in Fourier space
\be
\langle a_{\kk_1} \cdots a_{\kk_n} \rangle_c
\ee
Under the assumption of statistical homogeneity and isotropy, this correlation function vanishes unless $\kk_1 \cdots \kk_n$ form a polygon. This polygon may be parametrised by the lengths of its sides $k_1 \cdots k_n$ and by diagonals $k^d_1 \cdots k^d_m$ needed to fix the shape via a chosen triangulation. The correlation function of order $n$ thus has $2n-3$ degrees of freedom in 2D (i.e. $m=n-3$), as can be seen on Fig.\ref{Fig:polygons2D} at \textcolor{blue}{fourth} (trispectrum) and \textcolor{red}{$n$-th} order.

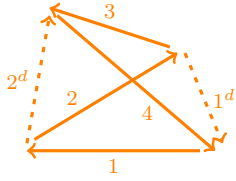
\begin{figure}
\begin{center}
\begin{tikzpicture}
\draw [->, very thick,orange] (2.3,-1) -- node[below] {1} (0,-1);
\draw [->, very thick,orange] (0.1,-0.85) -- (2,0.3);
\node[orange] at (0.6,-0.3) {2};
\draw [->, very thick,dash pattern=on 2pt off 3pt,orange] (2.1,0.3) -- node[right] {$1^d$} (2.6,-0.9);
\draw [->, very thick,orange] (1.9,0.38) -- node[above] {3} (0.3,0.9);
\draw [->, very thick,orange] (0.4,0.8) -- (2.5,-1);
\node[orange] at (1.6,-0.5) {4};
\draw [->, very thick,dash pattern=on 2pt off 3pt,orange] (0,-0.9) -- node[left] {$2^d$} (0.3,0.8);
\end{tikzpicture}
\caption{Parametrisation of the 3D trispectrum with two diagonal degrees of freedom in dashed lines.}
\label{Fig:polygons3D}
\end{center}
\end{figure}

Correspondingly, the correlation function of order $n$ has $3n-6$ degrees of freedom in 3D ($m=2n-6$), as can be seen on Fig.\ref{Fig:polygons3D} at \textcolor{orange}{fourth} order (trispectrum). 

The polyspectrum of order $n$, $\mathcal{P}^{(n)}(k_1 \cdots k_n, k^d_1 \cdots k^d_m)\,$, is then defined by~:
\ba
\nonumber \langle a_{\kk_1} \cdots a_{\kk_n} \rangle_c &= \int \frac{\dd^D k_1^d}{(2\pi)^D} \cdots \frac{\dd^D k_m^d}{(2\pi)^D} \;\mathcal{P}^{(n)}(k_1 \cdots k_n, k^d_1 \cdots k^d_m) \\
\label{Eq:defpolyspFourier} &\quad  \times \prod_g \, (2\pi)^D \, \delta\left(\kk_1(g)+\kk_2(g)+\kk_3(g)\right)
\ea
where $D$ is the dimension of the random field ($D=3$ for the galaxy distribution), and $g$ indexes the chosen triangulation of the polygon (e.g. for the 2D trispectrum $g=(1,2)$ and the triangulation is $(\kk_1,\kk_2,\kk_1^d);(-\kk_1^d,\kk_3,\kk_4)$).

Note that at orders $\geq 4$, polyspectra have more degrees of freedom in 3D than in 2D. However this does not happen at the bispectrum level which does not have diagonal degrees of freedom, as a triangle is flat and can be parametrised solely with its sides.\\
For some random fields (e.g. Gaussian or white-noise), the polyspectra may not depend on diagonal degrees of freedom, such polyspectra are called diagonal-independent and can be parametrised solely with the length of the sides $k_1 \cdots k_n$. In this case, Eq.\ref{Eq:defpolyspFourier} takes the simpler form~:
\be
\langle a_{\kk_1} \cdots a_{\kk_n} \rangle_c = (2\pi)^D  \, \delta(\kk_1 + \cdots + \kk_n) \times \mathcal{P}^{(n)}(k_{1\cdots n} )
\ee
\newline


The case of random fields on the sphere is similar to the 2D Fourier case, and can be defined simply with the replacements~:
\ba
\int \frac{\dd^2\kk^d}{(2\pi)^2} &\rightarrow \sum_{\ell^d m^d} \\
(2\pi)^2\delta(\kk_1+\kk_2+\kk_3) & \rightarrow G_{123}
\ea
the $n$-th order polyspectrum has $2n-3$ degrees of freedom and is defined through~: 
\ba
\nonumber \langle a_{\ell_1 m_1} \cdots a_{\ell_n m_n} \rangle_c &= \!\!\!\!\sum_{\substack{\ell^\mathrm{d}_1 \cdots \ell^\mathrm{d}_{n-3} \\ m^\mathrm{d}_1 \cdots \ell^\mathrm{d}_{n-3}}}\!\!\!\! \mathcal{P}^{(n)}(\ell_{1\cdots n}\, , \ell^\mathrm{d}_{1\cdots (n-3)})\\
\label{Eq:polyspecharmwdiag} & \qquad \quad \times \mathcal{G}_{1\cdots n}(\ell^\mathrm{d}_{1\cdots (n-3)},m^\mathrm{d}_{1\cdots (n-3)})
\ea
with
\begin{align}
\nonumber \mathcal{G}_{1\cdots n}(\ell^\mathrm{d}_{1\cdots (n-3)} , m^\mathrm{d}_{1\cdots (n-3)}) &= G_{1,2,1^d} \times G_{1^{d*}, 3, 2^d} \times \cdots \\
& \times G_{(n-4)^{d*} , n-2 , (n-3)^d} \times G_{(n-3)^{d*}, (n-1) , n}
\end{align}

The polyspectrum is diagonal-independent if $\mathcal{P}^{(n)}$ does 
not vary with $\ell^\mathrm{d}_1 \cdots \ell^\mathrm{d}_{n-3}$. In this case, 
Eq.\ref{Eq:polyspecharmwdiag} takes a simpler form :
\be
\langle a_{\ell_1 m_1} \cdots a_{\ell_n m_n} \rangle_c = \mathcal{P}^{(n)}(\ell_{1\cdots n}) \times \mathcal{G}_{1, \cdots , n}
\ee
with
\be
\mathcal{G}_{1, \cdots , n} = \int \dd^2\nn \; Y_{1 \cdots n}(\nn) = \sum_{\substack{\ell^\mathrm{d}_{1 \cdots (n-3)} \\ m^\mathrm{d}_{1 \cdots (n-3)}}} \mathcal{G}_{1\cdots n}(\ell^\mathrm{d}_{1\cdots (n-3)} , \, m^\mathrm{d}_{1\cdots (n-3)})
\ee


\section{Projection of 3D polyspectra on the sphere}\label{App:derivpolyspecirgen}
Noting for simplicity $g(z) = \frac{\dd r}{\dd z} \, a(z)$, we have :
\be
a_{\ell m} = i^\ell \int \frac{\dd^3\kk}{2\pi^2} \, \dd z \, g(z) \, j_\ell(k r) \, Y^*_{\ell m}(\hat{k}) \, j_\nu(\kk,z)
\ee
Hence the n-order correlation function takes the form :
\ba
\nonumber
\langle a_{\ell_1 m_1} \cdots a_{\ell_n m_n} \rangle = {} & i^{\ell_1+\cdots+\ell_n} \!\!\int\! \frac{\dd^3\kk_{1\cdots n}}{(2\pi^2)^n} \, \dd z_{1\cdots n} \Big[g(z_i)\, j_{\ell_i}(k_i r_i) \\
\nonumber & \qquad\qquad \times Y^*_{\ell_i m_i}(\hat{k}_i) \Big]_{i=1\cdots n} \\
\label{Eq:proj3Dpolysp-general} & \qquad\qquad \times \!\!\!\!\!\!\!\!\!\!\!\!\!\!\underbrace{\langle \left[\texttt{j}_\nu(\kk_i , z_i )\right]_{i=1\cdots n} \rangle}_{ (2\pi)^3 \, \mathcal{P}^{(n)}_\texttt{j}(\kk_{1 \cdots n},z_{1 \cdots n}) \, \delta^{(3)}(\kk_1 + \cdots + \kk_n)}\\
\nonumber = {} & \frac{(2\pi)^3}{(2\pi^2)^n} \, i^{\ell_1+\cdots+\ell_n} \!\!\int\! k^2_{1 \cdots n} \dd k_{1 \cdots n} \, \dd z_{1\cdots n} \\
\nonumber & \quad \times \left[g(z_i)\, j_{\ell_i}(k_i r_i)\right]_{i} \mathcal{P}^{(n)}_{j}(\kk_{1 \cdots n},z_{1 \cdots n}) \\
& \times \underbrace{\int \dd^2 \hat{k}_{1 \cdots n} \left[Y^*_{\ell_i m_i}(\hat{k}_i)\right]_{i} \delta^{(3)}(\kk_1 + \cdots + \kk_n) }_{\equiv A^{(n)}(\ell_{1 \cdots n}, m_{1 \cdots n}, \kk_{1 \cdots n})}
\ea
if $\mathcal{P}^{(n)}_{j}$ is diagonal-independent, and with
\ba
\nonumber A^{(n)}(\ell_{1 \cdots n}, m_{1 \cdots n}, \kk_{1 \cdots n}) = {} & \int \dd^2 \hat{k}_{1 \cdots n} \frac{\dd^3 \mathbf{x}}{(2\pi)^3}\\
\label{Eq:Anline1} & \times \left[Y^*_{\ell_i m_i}(\hat{k}_i)\right]_{i} e^{i(\kk_1 + \cdots + \kk_n)\cdot \mathbf{x}}\\
\nonumber = {} & \frac{(4\pi)^n}{(2\pi)^3} \int \dd^2 \hat{k}_{1 \cdots n} \,\dd^3 \mathbf{x} \\
\nonumber & \times \sum_{\ell'_{1 \cdots n} m'_{1 \cdots n}}\!\!\!\! i^{\ell'_1+\cdots+\ell'_n} \Big[Y^*_{\ell_i m_i}(\hat{k}_i) \\
\label{Eq:Anline2} & \times Y_{\ell'_i m'_i}(\hat{k}_i) \, j_{\ell'_i}(k_i x) \, Y^*_{\ell'_i m'_i}(\hat{x})\Big]_{i} \\
\nonumber = {} & \frac{(4\pi)^n}{(2\pi)^3} i^{\ell_1+\cdots+\ell_n} \!\!\! \int \! x^2 \dd x \left[ j_{\ell_i}(k_i x) \right]_{i} \\
\label{Eq:Anline3} & \times \underbrace{\int \dd^2 \nn \left[ Y^*_{\ell_i m_i}(\nn) \right]_{i=1\cdots n}}_{\equiv \,\mathcal{G}_{1\cdots n}}
\ea
where we introduced the Fourier form of the Dirac in line \ref{Eq:Anline1}, the Rayleigh expansion of $e^{i\kk_i\cdot\xx}$ in line \ref{Eq:Anline2}, used the orthonormality of the spherical harmonics and the definition of the generalised Gaunt coefficient in line \ref{Eq:Anline3} and the fact that it is a real number.\\
Hence the n-order (diagonal-independent) polyspectrum is :
\ba
\nonumber \mathcal{P}^{(n)}_\mathrm{IR}(\ell_1 , \cdots , \ell_n) = {} & \left(\frac{2}{\pi}\right)^n (-1)^{\ell_1+\cdots+\ell_n} \!\!\int\! k^2_{1 \cdots n} \dd k_{1 \cdots n} \, \dd z_{1\cdots n} \, x^2 \dd x \\
& \left[g(z_i)\, j_{\ell_i}(k_i r_i) j_{\ell_i}(k_i x) \right]_{i} \mathcal{P}^{(n)}_{j}(\kk_{1 \cdots n},z_{1 \cdots n})\\
\nonumber = {} & \left(\frac{2}{\pi}\right)^n (-1)^{\ell_1+\cdots+\ell_n} \!\!\int\! k^2_{1 \cdots n} \dd k_{1 \cdots n} \, \dd z_{1\cdots n} \, x^2 \dd x \\
\nonumber & \left[a(z_i)\, \left.\frac{\dd r}{\dd z}\right|_{z_i} \, j_{\ell_i}(k_i r_i) \, j_{\ell_i}(k_i x) \, \overline{j}(\nu,z_i)\right]_{i} \\
\label{Eq:Pn_IR_b4limber} & \quad \times \mathcal{P}^{(n)}_\mathrm{gal}(\kk_{1 \cdots n},z_{1 \cdots n})
\ea
where, for the last line, we used $\mathcal{P}^{(n)}_{j} = \overline{j}^n \, \mathcal{P}^{(n)}_\mathrm{gal}$ (valid for non shot-noise terms, with the flux-abundance independence assumption).\\
We can assume that, as a function of $k_i$, $\mathcal{P}^{(n)}_\mathrm{gal}$ varies slowly compared to the bessel functions oscillations (the so-called Limber approximation). Then we have
\ba
\nonumber \int\! k^2_{1 \cdots n} \dd k_{1 \cdots n} \left[j_{\ell_i}(k_i r_i) \, j_{\ell_i}(k_i x)\right]_{i} \mathcal{P}^{(n)}_\mathrm{gal}(\kk_{1 \cdots n},z_{1 \cdots n}) \qquad \\
\approx \mathcal{P}^{(n)}_\mathrm{gal}(k^*_{1 \cdots n},z_{1 \cdots n}) \underbrace{\int\! k^2_{1 \cdots n} \dd k_{1 \cdots n} \left[j_{\ell_i}(k_i r_i) \, j_{\ell_i}(k_i x)\right]_{i}}_{=\left[\frac{\pi}{2 x^2} \, \delta(r_i-x)\right]_{i=1\cdots n}}
\ea
where $k^*=(\ell+1/2)/r$ is the peak of the bessel function.\\
And Eq.\ref{Eq:Pn_IR_b4limber} simplifies to :
\be
\mathcal{P}^{(n)}_\mathrm{IR}(\ell_1 , \cdots , \ell_n) = \int \frac{r^2 \dd r}{r^{2n}} a^n(z) \, \overline{j}^n(\nu,z) \, \mathcal{P}^{(n)}_\mathrm{gal}(k^*_{1 \cdots n},z)
\ee
with $k^*=(\ell+1/2)/r$ , r=r(z), and because of parity invariance $\ell_1+\cdots+\ell_n$ is even (otherwise $\mathcal{P}^{(n)}$ is zero).

In particular at order 3, we get the bispectrum :
\ba
\nonumber b_{\ell_1 \ell_2 \ell_3} = {} & \left(\frac{2}{\pi}\right)^3 (-1)^{\ell_1+\ell_2+\ell_3} \int k^2 _{123} \dd k_{123} \, \dd z_{123} \, x^2 \dd x \\
\nonumber & \left[ a(z_i) \, \left.\frac{\dd r}{\dd z}\right|_{z_i} \, \overline{j}(\nu,z_i) \, j_{\ell_i}(k_i r_i) \, j_{\ell_i}(k_i x) \right]_{i=123} \\
& \qquad \times \, B_\mathrm{gal}(k_{123}, z_{123})
\ea
And Limber approximation simplifies it to~:
\be
 \bl = \int \frac{\dd z}{r^4} \,\frac{\dd r}{\dd z}\, a^3(z) \,\overline{j}^3(\nu,z)\, B_\mathrm{gal}(k^*_{123},z)
\ee
Again, except for shot-noise terms which are treated in Sect.\ref{Sect:shot-noise}.


\bibliographystyle{mn2e}
\bibliography{bibliography}

\end{document}